\documentclass[twocolumn]{article}
\usepackage{cite}
\usepackage{amsmath,amssymb,amsfonts}
\usepackage{graphicx}
\usepackage{textcomp}
\usepackage{multicol}
\usepackage{multirow}
\newcommand{\comment}[1]{}
\usepackage{xr}

\usepackage{titling}

\usepackage[a4paper, total={7.05in, 10.1in}]{geometry}

\usepackage{xcolor}

\usepackage{url}

\usepackage{algorithm}% http://ctan.org/pkg/algorithm
\usepackage{algpseudocode}% http://ctan.org/pkg/algorithmicx

\usepackage{marginnote}

\newcommand{\revb}[1]{#1} % red
\newcommand{\revc}[1]{#1} % blue

\newcommand{\rest}[1]{#1} % blue

\begin{document}

\color{black}{

\title{Sampling possible reconstructions of undersampled acquisitions in MR imaging with a deep learned prior}
\author{Kerem C. Tezcan$^\dagger$, Neerav Karani$^\dagger$, Christian F. Baumgartner$^\dagger {}^\ddag$ and Ender Konukoglu$^\dagger$\thanks{Accepted to IEEE Transactions in Medical Imaging for publication.
This work was supported by the Swiss National Science Foundation under Grant 205321\_173016. Emails are \{tezcan, nkarani, kender\}@vision.ee.ethz.ch and christian.baumgartner@uni-tuebingen.de} \\ \\ $^\dagger$ Computer Vision Lab, ETH Zürich, Switzerland \\ $^\ddag$ Medical Image Analysis Group, University of Tübingen, Germany}

\maketitle

\begin{abstract}
Undersampling the k-space during MR acquisitions saves time, however results in an ill-posed inversion problem, leading to an infinite set of images as possible solutions. Traditionally, this is tackled as a reconstruction problem by searching for a single "best" image out of this solution set according to some chosen regularization or prior. This approach, however, misses the possibility of other solutions and hence ignores the uncertainty in the inversion process. In this paper, we propose a method that instead returns multiple images which are possible under the acquisition model and the chosen prior to capture the uncertainty in the inversion process. To this end, we introduce a low dimensional latent space and model the posterior distribution of the latent vectors given the acquisition data in k-space, from which we can sample in the latent space and obtain the corresponding images. We use a variational autoencoder for the latent model and the Metropolis adjusted Langevin algorithm for the sampling. We evaluate our method on two datasets; with images from the Human Connectome Project and in-house measured multi-coil images. We compare to five alternative methods. \rest{Results indicate that the proposed method produces images that match the measured k-space data better than the alternatives, while showing realistic structural variability. Furthermore, in contrast to the compared methods, the proposed method yields higher uncertainty in the undersampled phase encoding direction, as expected.}


% \rest{Results indicate that the proposed method is capable of producing different images that match the ground truth in regions where acquired k-space data is informative, while showing realistic structural variations in regions where acquired k-space data is not informative.}
%which overlap with the ground truth for regions where k-space data is informative enough, and can capture the possibility of different reconstructions by showing realistic structural variations in regions where the k-space data is not informative enough. 

\end{abstract}

\textcolor{black}{}
\section{Introduction}
Undersampling the k-space in MR imaging reduces scan time by speeding up acquisition, allowing a higher throughput as well as higher comfort for patients. However, contrary to a fully acquired k-space, where an inverse Fourier transform is mostly enough to uniquely determine the underlying image up to measurement noise, the undersampled acquisition leads to an underdetermined system of equations. Mathematically, this means that there are infinitely many images that match the acquired portion of the k-space data and it is impossible to know which one is the underlying image in a general setting. 

Traditionally, this problem of infinitely many solutions has been tackled as a deterministic reconstruction problem, where different methods were proposed to choose a single "best" image out of the set of possible images. \rest{This was achieved by introducing prior information in addition to the data consistency term through defining a regularization term that prefers a solution according to properties} such as smoothness~\cite{l2reg}, sparsity~\cite{Lustig2007} or agreement with calibration data~\cite{tsao}. This converts the problem into a well-posed regularized inverse problem and allows to obtain a single solution as the reconstructed image. Another way of seeing this approach is from the Bayesian framework, where the data consistency and regularization terms correspond to the data likelihood and prior terms, respectively. Here, again, a suitable prior along with the data likelihood allows to write the corresponding posterior probability which then can be maximized to obtain the maximum-a-posteriori (MAP) image as the single 
``best'' reconstruction~\cite{tezcan}.

By providing a single reconstructed image as their output, these approaches miss the uncertainty in the solution that arises due to the missing portion of the k-space data. The reconstructed image is formed using information from the measured data and the prior, where the measured data matches the underlying true image up to the measurement noise while the prior \rest{compensates for} the unmeasured information. However, the image that best satisfies the prior need not be the same as the true image.  %Furthermore, in case the reconstructed image differs from the underlying true image, there is no possibility to extract this knowledge from only the reconstructed image itself.

An alternative approach, which we will pursue in this work, is producing multiple images as solutions to the ill-posed inverse problem, where the images match the measured data while being highly likely according to the prior at the same time. Such an approach is capable of capturing the uncertainty in the inversion due to the missing data.

As the proposed approach provides multiple images, clinical, research or further analysis tasks \rest{(e.g. segmentation for which we show some examples)} can be applied separately on each image to propagate the uncertainty to any given task's output. Alternatively the uncertainty can be quantified at the image level and passed on to the following tasks, such as in the form of a mean image along with a standard deviation map. Inspecting the images or the quantified uncertainty for different regions can also be indicative of which parts of the images might be more prone to differ from the underlying true image. To the best of our knowledge, this is the first time such an approach is being proposed for undersampled MRI acquisition.

In recent years, though not directly related to the uncertainty due to missing data in undersampled MRI, there have been research efforts to quantify uncertainty in different aspects of the image reconstruction problem, especially with the deep learning based models~\cite{gal_whatuncertainties}. One such aspect is the \emph{epistemic} or model uncertainty, i.e. the uncertainty in the mapping learned by the neural network used in the reconstruction, which can be obtained using approaches such as drop-out~\cite{gal_dropout}. However, these approaches quantify the uncertainty due to the ambiguity in the network parameters and do not capture the uncertainty due to the missing k-space data. Epistemic uncertainty tends to zero as the training set size goes to infinity~\cite{tanno_uncert}. However, increasing training samples cannot be expected to diminish the ambiguity due to missing k-space data. Another aspect is the so-called \emph{aleatoric} uncertainty due to noise or other ambiguities in the images or the labels, which is more relevant to reconstruction from undersampled MRI. Quantifying aleatoric uncertainty can be approached using, for instance, heteroscedastic models. These models predict a different variance value for each output pixel and hence can in principle learn to predict high variance for pixels where the model expects to have incorrect mean predictions~\cite{schlemper_heterosced_mri}. There is one important limitation to these models. They generate only second order statistics of pixel-wise marginal distributions. \rest{While such a model can predict where individual pixels may be inaccurate,} it cannot propose different possible reconstructions. For this, \rest{it is crucial} to capture the non-Gaussian pixel-wise distributions and non-trivial statistical dependencies across different pixels. Hence, these models are limited in the information they can provide~\cite{schlemper_heterosced_mri}. Being able to generate samples makes uncertainty propagation trivial for any following task, while only generating second order statistics makes this non-trivial. \revc{Furthermore, the g-factor~\cite{sense} as a conventional error measure for multi-coil reconstruction estimates the noise amplification relying on the coil geometry and the undersampling pattern and does not capture the uncertainty due to the used prior.} %not being able to generate samples means not allowing the possibility to quantify uncertainty for following tasks performed on the images.
% \kct{does this discussion about the heterosced. models make sense?}

One work that has similar aims as this paper is by Adler et al.~\cite{adler_cwgan}. In this work, the authors train a modified conditional Wasserstein generative adversarial network (\mbox{cWGAN}) that generates high dose counterparts of CT images from low-dose measurements. However, this approach lacks a data likelihood term and thus, does not explicitly model the known physics or measurement noise of the imaging procedure. As such, there are no guarantees that the samples will be from the true posterior. Instead of using an explicit physics-based imaging model, the cWGAN requires supervised training with pairs of undersampled-fully sampled images. Hence, a separate cWGAN has to be trained for all different undersampling schemes and factors for best performance. Furthermore, though the authors modify their discriminator to reduce the mode collapse associated with the WGAN, they do not completely avoid it, leading to a possibly poorer implicit prior. Lastly, the model aims to minimize the Wasserstein distance between the predicted posterior and the true posterior. However, it is often impossible to have varying samples from the true posterior, in reality only one fully measured image is available for each low-dose measurement image. Therefore, the available training samples may not be able to support an accurate posterior approximation and minimizing the Wasserstein distance may converge to a degenerate version of the true posterior. 

In this paper, we identify two ideas that motivate us in proposing a new method that overcomes the limitations of the works mentioned above. The first one is regarding the geometry of the space of MR images. We make the assumption that the MR images live around a low dimensional subspace in the high dimensional image space and that we can learn a mapping from a low dimensional latent space to this subspace. This assumption has been demonstrated empirically in our prior work~\cite{tezcan}. Then "walking around" and sampling in the subspace of MR images can be simply implemented as walking around and taking samples in the latent space. Secondly, deep learning based data-driven priors have shown great value in inverse problems in general in recent years, as well as specifically in MR image reconstruction~\cite{gan_inverse, cascade_networks, tezcan}. Such methodology allows learning a powerful mapping between the latent space and the image space, facilitating the sampling. 

Embodying the ideas mentioned above, we propose a novel method based on a latent Bayesian model and Markov chain Monte Carlo (MCMC) sampling that addresses the issue of uncertainty in the inversion process. To this end, we use a variational autoencoder (VAE)~\cite{KingmaW13, rezendevae} trained on fully sampled MR images as our prior and utilize its lower dimensional latent space to do the sampling instead of sampling in the high dimensional image space. \rest{We set the posterior of the latent vectors given the measured k-space data as the target distribution of the MCMC to obtain the latent samples.} We use the Metropolis adjusted Langevin Algorithm (MALA)~\cite{girolami} as the MCMC method due to its effectiveness in high dimensional spaces. The latent samples coming from MALA are then guaranteed to be from the posterior and can be transformed to images using the decoder of the VAE and the measured k-space data. Although we use a VAE and MALA in our implementation, the framework is generic and can be used with other generative models as well as sampling schemes. We evaluate our method with data from the Human Connectome Project (HCP)~\cite{VANESSEN20122222} as well as in-house measured images~\cite{tezcan} for changing settings of undersampling ratios and measurement noise levels and compare it to other \rest{sampling and reconstruction} methods.

\section{Methods}~\label{sec_methods}
\subsection{Main idea and notations}\label{sec:main_idea}
Let us denote the MR images by $x\in\mathbb{C}^N$ and the observed undersampled noisy k-space data $y\in\mathbb{C}^{M c}$ with $c$ coils, (M$\leq$N). We model the acquisition process as $y=Ex+\eta$, where $E\in\mathbb{C}^{Mc \times N}$ is an \textit{extended} MR encoding operation and $\eta$ is  complex Gaussian noise in the k-space with $\eta\sim N(0,\Sigma_{ns})$ and $\Sigma_{ns}$ as the noise covariance matrix. Hence, the data likelihood term is given as $p(y|x) = N(y; Ex, \Sigma_{ns})$.
In our implementation, we apply pre-whitening~\cite{KlaasP.Pruessmann2001}, but keep $\Sigma_{ns}$ in the equations for clarity. The extended encoding operation is defined as
\begin{equation}
E = \tilde{E}B\varphi P s,
\end{equation}
based on the usual encoding matrix $\tilde{E} = UFS$, with $U$ as the undersampling and $F$ as the Fourier operations and $S$ as the coil sensitivities. Further, the matrices $B$, $\varphi$ and $P$, are used to model the effect of the bias field, the MAP phase and to implement padding/cropping, respectively. Finally, $s$ is a scalar that introduces an invariance to scaling between the samples. We describe these in more detail in the Appendix due to space restrictions.

The goal of our work is to sample from the posterior distribution of images given the observed undersampled noisy k-space data, $p(x|y)$. \rest{Note that this approach is in contrast to the more conventional image reconstruction, where one aims to find a single image from this posterior as the reconstructed image, whereas our objective here is to characterize the whole posterior through multiple samples. }

\rest{One approach for obtaining such samples is to sample from $p(x|y)$ directly in the image space.} 
\rest{However, the high dimensionality of the image space} renders simple sampling methods \rest{(e.g. rejection sampling or vanilla MCMC~\cite{bishop})} very inefficient \rest{- they would need too many samples to adequately explore the image space to generate a good representation of the posterior, especially when assuming that high probability regions in the image space are not dense but rather form a lower dimensional subspace.}

\rest{To overcome this difficulty, we consider a latent variable model, where samples from a low-dimensional latent space generate the high-dimensional images. We write this as 
\begin{equation}\label{eqn:initial}
    p(x|y) = \int p(x,z|y) dz = \int p(x|z,y) p(z|y) dz,
\end{equation}
with z as the latent variable which is far lower dimensional than the image space. With this formulation we can use ancestral sampling, i.e. i) first sample $z^t$s from $p(z|y)$ and ii) then sample $x^t$s from $p(x|z^t, y)$. Here the intuition is that the samples $z^t$ are latent samples close to the measured k-space data and hence $x^t$ are as well. The main advantage of this two step process is that the latent space has fewer dimensions and is more densely populated than the image space, hence the proposed sampling is more efficient. Furthermore, as will become evident later, both terms $p(x|z,y)$ and $p(z|y)$ require a generative model, $p(x|z)$ with a simple prior $p(z)$ for which we use a variational autoencoder (VAE).}

\rest{The rest of this section is organized as follows. First, we describe the key ideas of our method: i) procedure of sampling from the posterior $p(z|y)$ in the latent space (Sec.~\ref{sec:sampling_lat_space}), ii) how we form this posterior in Sec.~\ref{sec:posterior} and iii) the procedure for converting the latent space samples to image samples (Sec.~\ref{sec_sampling_image_space}). Next, we give more detailed descriptions to i) the modifications in our VAE with respect to a vanilla VAE in Sec.~\ref{sec:VAE}, and finally, ii) how we calculate the matrix inversions arising in the components.}

\begin{algorithm}
\caption{l-MALA sampling}
\begin{algorithmic}[1]
\item Undersampled k-space data $y$, encoding operation $E$, trained VAE, MAP reconstruction $x_{MAP}$  
\Procedure{l-MALA}{$y$, $E$, VAE, $x_{MAP}$}
    \State \text{\# 0. Get the initial latent vector from the MAP}
    \For{$t=0:T-1$}
        \State \text{\# 1. First create the target distr. in TensorFlow}
        \State $\gamma^*  \gets \min_\gamma || (\Sigma_{x}^{-1}+E^H\Sigma_{ns}^{-1}E)\gamma - \Sigma_{x}^{-1}\mu_x ||_2^2$
        \State $\log p(y|z) \gets \mu_x^H\Sigma_{x}^{-1}\gamma^* + 2\text{Re}\left\{y^H\Sigma_{ns}^{-1}E\gamma^*\right\} 
     - \frac{1}{2} \mu_x^H\Sigma_x^{-1}\mu_x$
     
        \State \text{\# 2. Now take a step in the latent space}
        \State $\zeta \sim N(0,1)$
        \State $\hat{z}^{t+1} = z^t + \tau \nabla_z \log p(z|y)|_{z=z^t} + \sqrt{2\tau}\zeta $
        
        \State \text{\# 3. Accept or reject this}
        \State $\alpha \gets \min \left\{ 0 , \log \left[\frac{p(\hat{z}^{t+1}|y) q(z^t|\hat{z}^{t+1})}{p(z^t|y) q(\hat{z}^{t+1}\mid z^t)} \right]\right\}$
        \State $u \sim \log (\text{Unif}(0,1))$
        \If{$\alpha > u$} 
            \State $z^{t+1} \gets \hat{z}^{t+1}$
        \Else
            \State $z^{t+1} \gets z^{t}$
        \EndIf
        
        \State \text{\# 4. Convert the latent sample to an image}
        \State $\gamma^*  \gets \min_\gamma || (E_F\Sigma_x^{-1}E_F^H + U^H\Sigma_{ns}^{-1}U)\gamma - (E_F\Sigma_x^{-1}E_F^H\mu_x(z^{t+1}) + U^H\Sigma_{ns}^{-1}y) ||_2^2$\
        \State $x^t \gets E_F^H \gamma^*$
        
    \EndFor
\EndProcedure
\end{algorithmic}
\label{alg:sampling}
\end{algorithm}

\subsection{Sampling in the latent space}\label{sec:sampling_lat_space}

\rest{Here we assume we can calculate the target posterior $p(z|y)$. This posterior provides latent samples that match the measured k-space data and thus, the images associated with these latent samples will match the k-space data too. We also assume we can calculate its gradient $\nabla_z \log p(z|y)$, as well. In this section we only describe the sampling procedure and we will describe how to calculate these terms in Sec.~\ref{sec:posterior}}.

\rest{We use the Metropolis adjusted Langevin algorithm (MALA)~\cite{girolami} to sample from $p(z|y)$. MALA is a variant of Markov chain Monte Carlo. It consists of (a) a random walk given by Langevin dynamics and (b) an acceptance scheme following the Metropolis-Hastings algorithm.} The random walk for MALA with the target distribution $p(z|y)$ is written as: 
\begin{equation}~\label{eqn_random_walk}
    \hat{z}^{t+1} = z^t + \tau \nabla_z \log p(z|y)|_{z=z^t} + \sqrt{2\tau}\zeta,\ \ \zeta\sim N(0,1)
\end{equation}
where $\tau$ is the step size.
As can be observed, the update step for the random walk is composed of two terms:
\rest{(1) The first term, $\tau \nabla_z \log p(z|y)$, pulls the random walk towards high probability regions of the posterior, preventing it from moving far away from such regions.
(2) The second term models the randomness with a Gaussian distributed variable $\zeta$, and encourages the random walk to explore the latent space rather than only converging to the maximum of the posterior.}
Further, the discrete nature of the walk requires a Metropolis-Hastings correction to be applied to ensure convergence.
This means that a sample is accepted as $z^{t+1} = \hat{z}^{t+1}$ with log probability  $\alpha = \min \left\{ 0 , \log \left[\frac{p(\hat{z}^{t+1}|y) q(z^t|\hat{z}^{t+1})}{p(z^t|y) q(\hat{z}^{t+1}\mid z^t)} \right]\right\}$, otherwise $z^{t+1} = z^{t}$.
Here, $q$ is the proposal distribution, and is given as $q(z'\mid z) \propto \exp \left( - \frac{1}{4 \tau} \| z' - z - \tau \nabla \log p(z|y) \|_2^2 \right)$.
Finally, in order to avoid a long burn-in period, we can initialize the chain close to the mode of the posterior instead of starting with a randomly chosen $z^0$. This can be achieved by encoding the MAP image into the latent space and using its mean as $z^0$.

\subsection{Obtaining the posterior \rest{in the latent space,} $p(z|y)$}\label{sec:posterior}

\rest{So far, we have described the procedure of sampling in the latent space (Sec.~\ref{sec:sampling_lat_space}). In this section, we first briefly describe the generative model used for the latent space, i.e. the VAE, and then describe the method to construct the posterior $p(z|y)$ from which we sampled in Sec.~\ref{sec:sampling_lat_space}.}

\subsubsection{\rest{Latent variable prior model}}\label{sec:vae}
\rest{We use a VAE~\cite{KingmaW13, rezendevae} as the prior model of MR images and train it using fully sampled images. The VAE consists of an encoder, $q(z|x) = N(\mu_z(x),\Sigma_z(x))$, and a decoder, $p(x|z) = N(\mu_x(z),\Sigma_x)$, both parameterized with neural networks. $z\in \mathcal{R}^D$ (D$<<$N) is the latent representation, distributed according to a Gaussian prior $p(z) = N(\mu_{pr}, \Sigma_{pr})$. We use a diagonal non-isotropic covariance matrix for $\Sigma_z$, an isotropic diagonal matrix for $\Sigma_x$ with a fixed value and a block diagonal for the $\Sigma_{pr}$. From now on, we drop the $(x)$ and $(z)$ notation, i.e. write $\mu_x \triangleq \mu_x(z)$ unless necessary. The VAE is trained to maximize the evidence lower bound (ELBO), which approximates the log likelihood, $\log \:p(x)$. For the prior, $p(z)$, we empirically estimate the parameters $\mu_{pr}$ and $\Sigma_{pr}$ from training data~\cite{anna_empprior}. This step is different from a vanilla VAE, and explained in more detail in Sec.~\ref{sec:VAE}. }

\subsubsection{\rest{Computing the unnormalized posterior}}
\color{black}
\rest{Note that the derivative in random walk update step (Eqn.~\ref{eqn_random_walk}) as well as the acceptance term described in Sec.~\ref{sec:sampling_lat_space} can be computed using the unnormalized posterior distribution of $z$: $p(z)p(y|z) \propto p(z|y)$, that is, the normalization constant $p(y)$ does not appear in the equations.
The unnormalized posterior, $p(z)p(y|z)$, is a product of two terms:
(1) For the first term, we use the empirical prior, $p(z)=N(\mu_{pr}, \Sigma_{pr})$.
(2) We write the second term as a marginalization over images $x$:
\begin{equation}\label{eq:marg}
    p(y|z) = \int p(y,x|z) dx = \int p(y|x) p(x|z) dx,
\end{equation}
where we use the VAE decoder as $p(x|z)$ and the conditional independence assumption $p(y|x,z) = p(y|x)$.
In this formulation, the images function as intermediate variables - they connect the latent space representations with the k-space data.}
% This expression can be interpreted as two terms glued together with the images functioning as the intermediate variables, connecting the latent space with the k-space.
This integral can be evaluated analytically to yield another Gaussian distribution~\cite{katarina_marg}.
After isolating the terms that are constant with respect to $z$ and taking the logarithm, this distribution is given as: 
\begin{align}  
     \log p&(y|z)  = \mu_x^H\Sigma_{x}^{-1}(\Sigma_{x}^{-1}+E^H\Sigma_{ns}^{-1}E)^{-1}\Sigma_{x}^{-1}\mu_x \nonumber \\
     &+2 \text{Re}\left\{ y^H\Sigma_{ns}^{-1}E(\Sigma_{x}^{-1}+E^H\Sigma_{ns}^{-1}E)^{-1}\Sigma_{x}^{-1}\mu_x\right\}  \label{eq:marg2} \\
     &- \frac{1}{2} \mu_x^H\Sigma_x^{-1}\mu_x +C. \nonumber
\end{align}
where C is a constant with respect to $z$ and $^H$ denotes the complex conjugate transpose. Please refer to Appendix for details of this derivation. Also, notice that we need to calculate the inverse the matrix $(\Sigma_{x}^{-1}+E^H\Sigma_{ns}^{-1}E)^{-1}$. We describe how to do this in Sec.~\ref{sec:comp_details}.

\comment{Finally, looking at the two terms in the posterior $p(z|y) \propto p(y|z)p(z)$ reveals more insights regarding the method. The first term drives the chain to regions where the $z^t$ values, when decoded as $p(x|z^t)$, lead to images which satisfy the data likelihood term $p(y|x)$ for $x$'s coming form the $p(x|z^t)$. On the other hand, the second term $p(z)$ tries to pull the chain towards the middle of the empirical Gaussian in the latent space, discouraging the chain to move away from the meaningful regions of the latent space. As such, a random walk in the latent space with this target distribution explores the areas which satisfy the data likelihood and the prior terms simultaneously.} 

\subsubsection{\rest{Interpreting terms of the unnormalized posterior}}
Considering the two terms in the unnormalized posterior, $p(y|z)p(z)$, reveals more insights regarding the method.
The first term $p(z)$ tries to pull the chain towards the middle of the empirical Gaussian in the latent space, discouraging the chain to move away from the meaningful regions of the latent space, where MR images reside.
On the other hand, the second term drives the chain to regions where the $z^t$ values, when decoded as $p(x|z^t)$, lead to images which satisfy the data likelihood term $p(y|x)$ for $x$'s coming form the $p(x|z^t)$.
As such, a random walk in the latent space with this target distribution explores the areas which satisfy the data likelihood and the prior terms simultaneously.

\subsection{\rest{Converting latent space samples to image samples}}\label{sec_sampling_image_space}

\revb{After obtaining the samples $\{z^t\}$ from the posterior $p(z|y)$ as described in Sec.~\ref{sec:sampling_lat_space}, we now need to convert them into image samples.
To this end, we sample from the posterior of images $x$, $p(x|z^t,y)$, given both the latent sample $z^t$ as well as the measured k-space data $y$ as motivated by the ancestral sampling procedure introduced in Sec.~\ref{sec:main_idea}. 
We define this posterior as:
\begin{equation}\label{eq:2ndstep}
    p(x|z,y)  = \frac{p(x,y|z)}{p(y|z)} \propto p(x|z) p(y|x);
\end{equation}}
\revb{Here, we again use the conditional independence assumption $p(y|x,z) = p(y|x)$. Notice that the terms $p(x|z)$ and $p(y|x)$ correspond to the decoder of the VAE and the data likelihood terms, respectively. 
As both terms on the right hand side of Eq.~\ref{eq:2ndstep} are Gaussian distributions, the posterior is also a Gaussian distribution: $p(x|z,y) = N(\mu_{x|z,y}, \Sigma_{x|z,y})$.
Further, we can derive a closed form solution for its mean given as
\begin{multline}\label{eqn:x_given_zy}
    \mu_{x|z,y} = E_F^H \left[ E_F\Sigma_x^{-1}E_F^H + U^H\Sigma_{ns}^{-1}U \right]^{-1} \\ \cdot\left[ E_F\Sigma_x^{-1}E_F^H\mu_x + U^H\Sigma_{ns}^{-1}y \right],
\end{multline}
\revb{where we defined $E_F$ as the encoding operation without the undersampling, i.e. $E = UE_F$. We present the derivation of the closed form solution in the Appendix for space considerations. Notice that we need to calculate the matrix inversion $\left[ E_F\Sigma_x^{-1}E_F^H + U^H\Sigma_{ns}U \right]^{-1}$, similar to the inversion in Eqn.~\ref{eq:marg2}. We describe how we do this in Sec.~\ref{sec:comp_details}.
For simplicity, we directly use the mean of this distribution as the image sample, that is $x^t = \mu_{x|z^t,y}$}.}

\rest{This concludes the sampling procedure, which we sum up in the Algorithm~\ref{alg:sampling}. In the following we describe how we define and calculate the building blocks used until now.}

\begin{figure}[t!]
    \centering
    \includegraphics[width=0.48\textwidth]{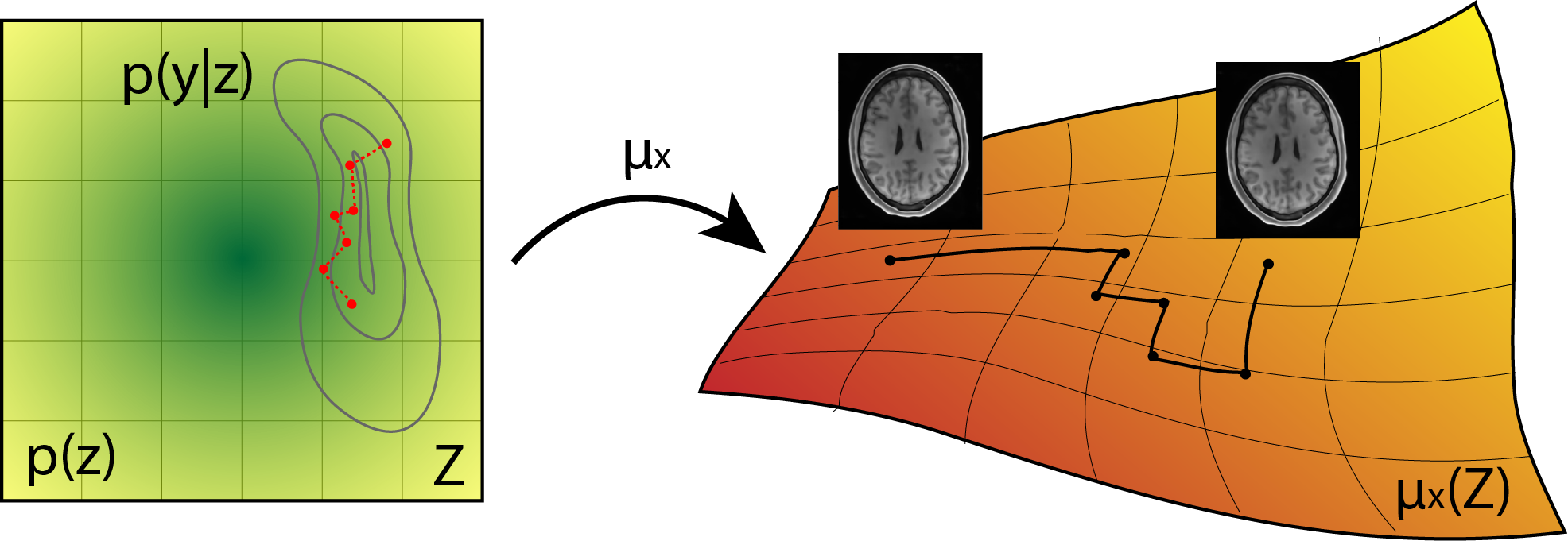}
    \caption{Left: illustration of sampling in the latent space, which is equipped with a prior shown in green. The contours show the regions where $z$'s lead to high data likelihood values for the measured k-space data. The random walk (red line) samples from the product of these two, i.e. the posterior $z^t\sim p(z|y)\propto(y|z)p(z)$. Right: illustration of sampling in the image space. The orange-yellow surface indicates the learned low-dimensional subspace around which MR images reside. \revb{Each sample $z^t$ in the latent space corresponds to a distribution in the image space, $p(x|z^t)$. We combine this with the measured data to obtain image samples, $x^t \sim p(x|y,z^t) \propto p(x|z^t)p(y|x)$.}}
    \label{fig:latentspace}
\end{figure}

\subsection{\rest{Computations details}}
\label{sec:comp_details}
\rest{In this section, we describe some computational details in computing the unnormalized posterior and using it to sample from the latent space (Sec.~\ref{sec:sampling_lat_space}).}

\rest{\textbf{(i) Gradient of the target distribution}:} For the random walk updates (Eqn.~\ref{eqn_random_walk}), we need the log-gradient of the target distribution. We write this as  $\nabla_z \log p(z|y) = \nabla_z \log p(y|z) + \nabla_z\log p(z)$. The $\nabla_z \log p(y|z)$ term can be obtained by automatic differentiation, as outputs of the decoder $\mu_x(z)$ and $\Sigma_x(z)$ are modeled as neural networks and thus, differentiable with respect to $z$. %Given that we can implement $p(y|z)$ in a software package that allows such differentiation. %which can be obtained by automatic differentiation as the outputs of the decoder $\mu_x$ and $\Sigma_x$ are differentiable w.r.t. z, since they are modeled by neural networks. 
Similarly, the prior $\nabla_z \log p(z)$ term is also straightforward and can be derived analytically. 

\rest{\textbf{(ii) Matrix inversion}:} For the sampling procedure, we need to evaluate the terms in Eq.~\ref{eq:marg2} at each iteration. This means the inverse of $(\Sigma_{x}^{-1}+E^H\Sigma_{ns}^{-1}E)$ has to be recomputed at each iteration, if the $\Sigma_x$ term depends on the z value. \rest{Even though we assume $\Sigma_x$ to be constant, inverting the matrix once and storing it is not possible due to memory issues (it is a $Nc\times Nc$ matrix)}. Further, approximating it as a diagonal matrix is also not an option as this would result in the loss of the aliasing information kept in the off-diagonals of the $E^H\Sigma_{ns}^{-1}E$ term. Similarly, Eq.~\ref{eqn:x_given_zy} used for sampling images from the latent samples also requires a similar inversion. Though this needs to be evaluated only once for fixed $\Sigma_x$, it is still too big for the memory. 

Instead we propose to use an iterative matrix inversion for these two terms that can also be applied when $\Sigma_x$ changes with $z$. To this end we write the inversions in a generic form as 
\begin{align}
        \gamma^*  = \min_\gamma || A\gamma - b ||_2^2, \label{eq:gammal2}
\end{align}
where $A$ is the matrix to be inverted and $b$ is the result vector. That is, we set for Eqn.~\ref{eq:marg2}, $A = \Sigma_{x}^{-1}+E^H\Sigma_{ns}^{-1}E$, $b = \Sigma_{x}^{-1}\mu_x$ and similarly, for Eqn.~\ref{eqn:x_given_zy}, $A = E_F\Sigma_x^{-1}E_F^H + U^H\Sigma_{ns}^{-1}U$, $b = E_F\Sigma_x^{-1}E_F^H\mu_x + U^H\Sigma_{ns}^{-1}y$.

We then solve this inversion as an optimization problem using conjugate gradients (CG) and obtain $\gamma^*$, which we plug-in to Eq.~\ref{eq:marg2} to yield
\begin{align}    
     \log p(y|z)  = \mu_x^H\Sigma_{x}^{-1}\gamma^* + 2\text{Re}\left\{y^H\Sigma_{ns}^{-1}E\gamma^*\right\} 
     - \frac{1}{2} \mu_x^H\Sigma_x^{-1}\mu_x,
\end{align}
where we dropped the constant C. Similarly, for Eqn.~\ref{eqn:x_given_zy} we then use
\begin{equation}
    \mu_{x|z,y} = E_F^H \gamma^*.
\end{equation}

Though we use CG here, other gradient based methods can be used as well, since the gradients are well defined. Furthermore fast Fourier transform (FFT) can be used in the operations $E$ and $E^H$, revoking the need to write the matrix explicitly and speeding up computations. 

\revc{The key advantage here is that all steps in the CG are differentiable with respect to z, so we can use automatic differentiation to differentiate $\gamma^*$ with respect to z, when the number of iterations ($N_\gamma$) for the CG is fixed. This allows us to take the gradients of $\log p(y|z)$ according to $z$. Notice this is very similar to the idea of "unrolled optimization", where an iterative process is written explicitly. Methods such as~\cite{hammernikvar, NIPS2016_6406} use this idea to calculate the gradients according to the network parameters, whereas here it is used to calculate the gradient according to the input vector.} Hence, before starting the Markov chain we select the parameter $N_\gamma$ for which we obtain a small L$_2$ error in Eq.~\ref{eq:gammal2} and keep this throughout the sampling. We observe that the error stays small throughout the sampling process for the chosen parameters for different $\mu_x$ values.

\subsection{Alternative ideas/shortcuts}
\subsubsection{Sampling from the decoder: $x^t \sim p(x|z^t)$} The most straight-forward way to sample images from a latent sample $z^t$ would be to sample directly from the decoder, i.e. $x \sim p(x|z^t)$, that is, to pass the $z^t$s through the VAE decoder and sample from $p(x|z)$. Similarly, for simplicity one could also take directly the mean of the decoder distribution as $x^t = \mu_x$. This, however, results in blurry images - a common problem in VAE generated samples. Furthermore, this takes the k-space data only implicitly into account while sampling $z^t$ but not while sampling $x^t$.

\subsubsection{Sampling from the encoder: $z^t \sim p(z|x_{MAP})$} \rest{An alternative to the proposed method is to use the approximation posterior, $q(z|x)$, modeled in VAE to do the sampling around a MAP estimate, $x_{MAP}$ or alternatively around the zero-filled image. This would entail (a) taking latent samples around the conditional distribution $z^t \sim q(z|x_{MAP})$ or $z^t \sim q(z|x_{ZF})$ and then (b) decoding these to the image space, similar to~\cite{pauly_uncertainty}. This approach is fundamentally different to the proposed method in that it takes the MAP image as the "true reconstruction" and samples only locally around it. Though these locally sampled images will still be in the subspace of MR images and show some structural variations, they are unlikely to explore the image space sufficiently to provide globally distinct images that are still coherent with the measured k-space data. Hence, such "local" sampling methods are inherently limited and cannot identify regions where the reconstruction has failed. In contrast, the proposed method can "globally" explore the latent space as long as the data likelihood is satisfied. Furthermore, $q(z|x)$ is only an approximate posterior distribution for a given $p(x|z)$ and $p(z)$, while we extract samples from the exact posterior. Finally, such a sampling scheme does not take the level of noise in the k-space into account, i.e. cannot guarantee that the sample diversity will increase with increasing k-space noise level. We also compare the two methods experimentally (refer to Sec.~\ref{sec:results} for the results).}

\subsection{\rest{Modifications in the VAE model: The 2D latent space with an empirical prior}}\label{sec:VAE}
\rest{The vanilla VAE model~\cite{KingmaW13} is described in the Appendix. In this subsection, we describe our modifications to the VAE, which allow us to model the distribution of full-sized MR images.}

\subsubsection{Fully-convolutional} \rest{Firstly, we make the VAE fully convolutional and use a 2D latent space of size $L_1$ x $L_2$ x D, which allows us to model full-sized images in the low dimensional subspace. Such a latent representation can be thought of as a latent image with D channels. Now, each spatial location in the latent image has a receptive field with respect to the input image. In order to minimize the overlap between the receptive fields of different pixels in the latent image, we employ high stride values in the convolutional layers. This allows us to adhere to the independence assumption of the latent pixels, and justify the use of a diagonal covariance matrix for $q(z|x)$.}

\subsubsection{Empirical prior in the latent space} \rest{In reality, contents in receptive fields corresponding to different spatial locations in a latent image are not entirely independent from each other, as there are global correlations in the image. To be able to model such correlations, we introduce an empirical prior as $p(z) = N(\mu_{pr}, \Sigma_{pr})$, similar to~\cite{anna_empprior}. 
This is the second modification with respect to vanilla VAEs. We compute the parameters of $N(\mu_{pr}, \Sigma_{pr})$ empirically using T samples $z_i \sim q(z|x_i)$ from T different training images $x_i$, after the VAE has been trained using a unit Gaussian prior.
The mean, $\mu_{pr}$, is calculated as the mean of the samples, $z_i$.}

\subsubsection{Estimating the covariance of the empirical prior} The estimation of a full covariance matrix, $\Sigma_{pr}$, is challenging due to i) its size and ii) large number of samples required for the estimation.
\rest{To tackle this issue, we first rank the latent channels in terms of their informativeness. To achieve this,} we apply a Kolmogorov-Smirnov test~\cite{kolmogorov} against the unit Gaussian separately for each latent channel to identify channels that are the least unit Gaussian in terms of the p-values, i.e. approximately the most informative.
Then, we form a combined block diagonal covariance matrix, where we calculate a full covariance matrix for the $K$ most informative (least Gaussian-like) channels of size $KL_1L_2$ x $KL_1L_2$.
We assume that the remaining latent channels are independent from each other and calculate only the spatial covariance matrix (of size $L_1L_2$ x $L_1L_2$) for each of them separately.
The proper combination of these block matrices yields the $\Sigma_{pr}$. \rest{This strategy allows us to reduce the number of samples, $T$, required for the estimation of $\Sigma_{pr}$.}
In practice, we do not form this matrix, but instead implement the operations as sparse matrix-vector multiplications. We set $K=10$, as our preliminary experiments showed that this covers the informative channels sufficiently. We also set $T$ high enough to ensure full rank of the estimated covariance matrix, finding $T=20000$ to be sufficient.

\section{Experimental Setup}

\subsection{Data, training details and compared methods}
We used T1 weighted slices from the full 3D volumes of 780 subjects from the HCP dataset~\cite{VANESSEN20122222} for training of the VAE. There were in total 202800 slices of size 252x308, with an isotropic resolution of $0.7\ mm$. We ran the N4 bias field correction on the images and used the corrected images for training. The training ran for 2250000 iterations (about 10 hours). We also trained another VAE after downsampling the images to 1mm isotropic resolution to work with lower resolution images for 1750000 iterations. For both, we augmented the images by translating them randomly (-4 to +4 pixels) in both directions and trained till convergence. 

For testing we used 9 axial slices from subjects in the HCP dataset, different than those used in training, without bias field correction. We additionally tested with 6 axial slices from in-house measured T1 weighted brain images of different subjects~\cite{tezcan}. These images have similar acquisition parameters as HCP and have an isotropic resolution of $1\ mm$. Furthermore these images are acquired with 13 coils and have non-zero phase.  We used ESPIRiT~\cite{espirit} to obtain the coil sensitivity maps, which we used in the MAP estimation and sampling.

For the experiments, we retrospectively apply Cartesian undersampling to the test images with different patterns for each image. We obtain these patterns by generating 100 different patterns and choosing the one with the highest peak-to-side ratio of the associated point spread functions. In all patterns the 15 central profiles are always sampled.

For comparison purposes, we also modified the code by Adler et al.~\cite{adler_cwgan} to work with magnitude of MR images and evaluated in our experimental setting (the authors provided private access to their repository for the code). This method requires supervised training, i.e. pairs of zero-filled and fully sampled images. We generated such a training set with zero-filled images by undersampling the training images, which were also used for the VAE. We used the images with their bias field. We generated an undersampling pattern as described above, seperately for each of the images. We trained multiple cWGANs fro different undersamplng ratios, for which we provide the details in the Appendix.  %For different undersampling ratios, we trained a separate cWGAN using a training set generated with that ratio.

We also trained a feed-forward heteroscedastic network as a baseline. This network has the same architecture as the VAE, without the KLD in the loss and outputs a pixelwise standard deviation alongside the mean prediction. It is trained for 4000000 iterations using the supervised training setup described for the cWGAN method.

Furthermore, we implement the local sampling method we described in Section~\ref{sec:posterior} for comparison purposes. For these we use the same VAE that is used in the proposed method. We take the samples around the MAP reconstruction.

\revb{Finally, we trained variational networks (Varnet)~\cite{hammernikvar}} (\url{https://github.com/visva89/VarNetRecon}) \revb{ for comparison purposes as a reconstruction baseline for different undersampling factors. The details of the network and its training are presented in the Appendix due to space restrictions.}

\subsection{Implementation Details}
We initialize the chain with the maximum-a-posteriori (MAP) images obtained by the deep density prior (DDP) reconstruction~\cite{tezcan} to avoid a long burn-in period and take 10000 samples in total, which takes around 2 hours for a slice with single coil data with our unoptimized implementation. We empirically determine the step size $\tau=4\times 10^{-4}$ to obtain an acceptance ratio around 0.3-0.5 and use the same for both the HCP images the in-house measured images. We take a lower number of samples (1000) for the cWGAN and local sampling methods as the effective sample size is also lower for the MCMC chain due to correlated samples.

We used Tensorflow~\cite{tensorflow} for the implementation of VAE related parts of the proposed method. The VAE is fully convolutional with all padded convolutions and has a 2 dimensional latent space with D=60 channels. We refer to the Appendix for the details of the architecture. For $\Sigma_x$ we use a diagonal matrix with equal diagonal values set at $0.02$. The same value was used for training of the VAE and sampling. 
%that was used for training of the VAE, i.e. 0.02.
$\Sigma_{ns}$ was estimated by taking the variance of a small region at the upper 150 pixels of the fully sampled k-space center in the undersampled data for all coils. In the implementation we pre-whitened the data and the sensitivity maps as described in~\cite{KlaasP.Pruessmann2001}. The number of iterations for the matrix inversion were determined empirically as $N_\gamma = 25$, which was enough to reduce the L2 error of the approximate inversion below 0.01\%.

For padding we use simple zero padding and cropping. For bias field estimation we used the N4 method~\cite{N4_paper} on the magnitude of the MAP estimation with default parameters. For the phase of the samples we took the MAP phase estimate. 

We trained a standard U-Net based segmentation network using the ground truth segmentations from the HCP dataset for the results shown in Figure 2. We omit the details of the segmentation networks here as irrelevant, expect emphasize that it is a deterministic network and hence any difference in the segmentation output is entirely coming from the variations in the input images.

\begin{figure}
    \centering
    \includegraphics[width=0.48\textwidth]{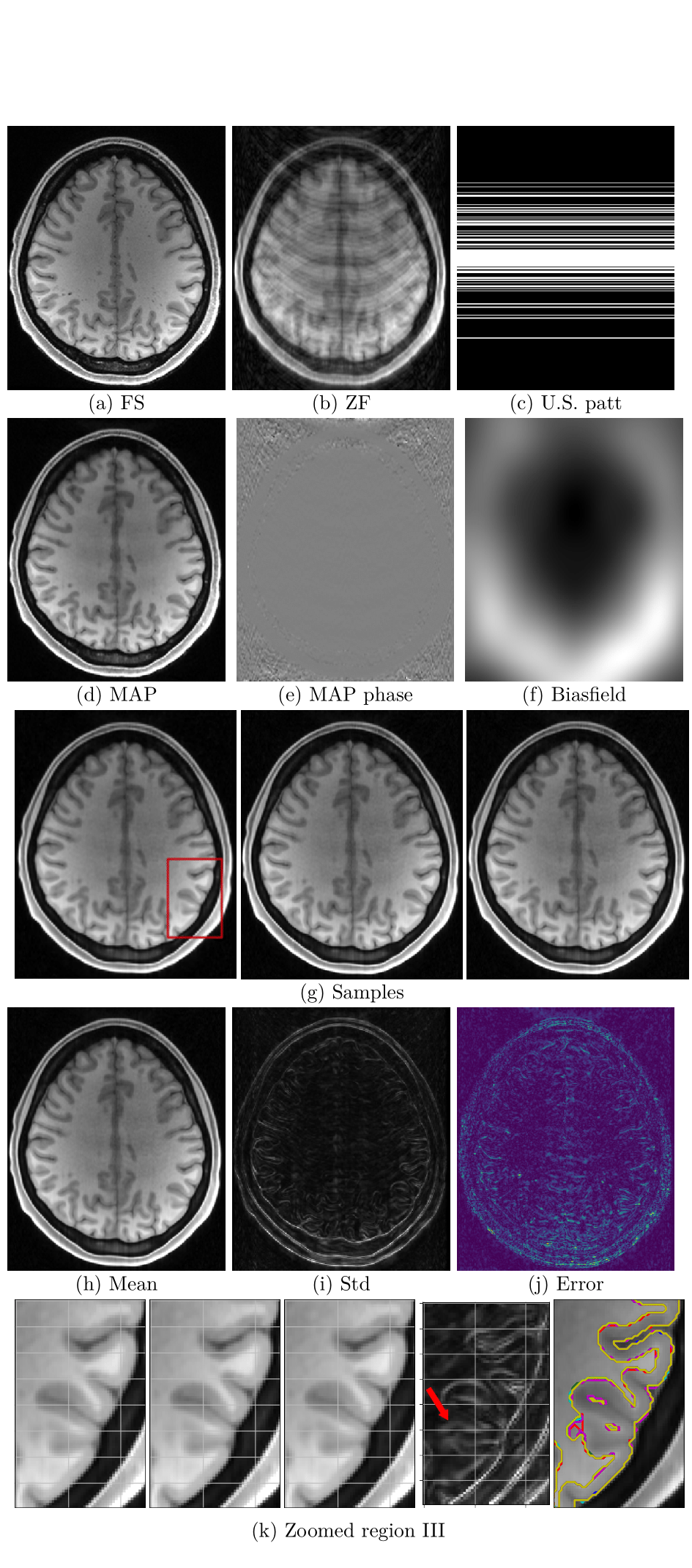}
    \caption{\revb{Results for the proposed latent MALA algorithm for R=5. FS, ZF and MAP denote the fully sampled, zero-filled and MAP estimation images. Phase image is scaled between $\pm \pi$. Third row presents three randomly chosen samples. (h-i) show the mean and pixelwise standard deviation (std) maps for all samples. (j) shows the absolute error map between the mean and the fully sampled image (clipped to 0-0.3). (k) shows three zoomed-in regions indicated in (g) for three different samples as well as the pixelwise std maps and contours of the gray matter segmentations from 10 different samples overlayed on the image. The grid lines are to aid visual inspection. As the variations are extremely difficult to see in this format, we strongly encourage the reader to look at the supplementary GIFs (\protect \url{https://github.com/kctezcan/sampling}).}}
    \label{fig:samples}
    %\vspace{-0.6cm}
\end{figure}

\section{Results}
\label{sec:results}
We show most of the visual results at high undersampling factors on purpose to make sure that the uncertainty in the inversion is high and demonstrate that the model is able to capture it. We present the results with the bias field put back in for convention although the method provides the samples bias free. We also multiply the images with their corresponding scale values to bring them to the same scale as the observed k-space data. The scale values stay mostly in the 0.95-1.1 range throughout the sampling procedure. We note that it is quite difficult to see the variations in the samples presented in this paper in print and highly encourage the reader to view the provided GIFs at \url{https://github.com/kctezcan/sampling}.

\begin{table*}[t]
\resizebox{\textwidth}{!}{
\begin{tabular}{|l|l|l|l|l|l|l|l|l|l|l|}
\hline
\multirow{2}{*}{Factor} & \multicolumn{5}{l|}{Absolute k-space error (x10$^{3}$)} & \multicolumn{5}{l|}{pairwise RMSE}                \\ \cline{2-11} 
                         & l-MALA & cWGAN & Local & VarNet & DDP & l-MALA & cWGAN & Local & VarNet & DDP \\ \hline
R=2 & 7.28 (1.47) & 22.70 (2.97) & 39.45 (4.49) & 17.73 (0.31) & 0.00 &               1.35 (0.52) & 6.9 (0.77) & 9.08 (0.73) & n.a. & n.a. \\ \hline
R=3 & 11.85 (1.49) & 33.64 (6.21) & 41.37 (4.86) & 18.14 (0.34) & 0.00&               1.50 (0.13) & 10.5 (1.49) & 8.53 (0.69) & n.a. & n.a. \\ \hline
R=4 & 15.50 (2.87) & 39.19 (5.02) & 42.83 (4.77) & 38.17 (5.11) & 0.00  &               2.27 (0.53) & 11.5 (1.29) & 8.08 (0.56) & n.a. & n.a. \\ \hline
R=5 & 17.03 (2.43) & 40.46 (4.73) & 45.22 (5.12) & 29.23 (2.96) & 0.00 &               2.16 (0.22) & 12.0 (1.43) & 8.03 (0.61) & n.a. & n.a. \\ \hline
\end{tabular}
}
\caption{\revb{Error metrics for the HCP test subjects at different undersampling factors. Absolute error is the summed absolute differences in the measured part of the k-space between the samples/reconstruction and the measured k-space data. Pairwise RMSE is the mean RMSE between sample image pairs and used here as a metric measuring sample diversity. Values shown in format: mean (std) for the sampling methods and the single value for the VarNet and DDP reconstruction methods. Presented values for abs. error are multiplied with $10^3$. }}
\label{tbl:hcp_abserr_pwrmse}
\end{table*}

\begin{table*}[t]
\resizebox{\textwidth}{!}{
\begin{tabular}{|l|l|l|l|l|l|l|l|l|l|l|l|l|l|l|l|}
\hline
\multirow{2}{*}{Factor} & \multicolumn{5}{l|}{NMSE (x10$^{3}$)} & \multicolumn{5}{l|}{pSNR}       & \multicolumn{5}{l|}{RMSE (\%)}             \\ \cline{2-16} 
                         & l-MALA & cWGAN & Local & VarNet & DDP & l-MALA & cWGAN & Local & VarNet & DDP & l-MALA & cWGAN & Local & VarNet & DDP \\ \hline
R=2 & 1.65 (0.79) & 4.55 (1.32) & 11.16 (3.10) & 1.49 (0.63) & 0.75 (0.42) &                40.49 (0.95) & 35.9 (0.92) & 10.48 (1.33) & 40.81 (0.76) & 44.03 (1.04) &  3.96 (0.87) & 6.7 (0.93) & 10.48 (1.33) & 3.80 (0.71) & 2.66 (0.68) \\ \hline
R=3 & 3.31 (1.38) & 9.71 (5.63) & 10.94 (3.03) & 1.98 (0.96) & 1.82 (0.89) &                37.36 (0.70) & 32.8 (1.50) & 10.37 (1.31) & 39.66 (0.92) & 40.05 (0.88) &  5.65 (1.05) & 9.6 (2.16) & 10.37 (1.31) & 4.35 (0.93) & 4.17 (0.91) \\ \hline
R=4 & 4.96 (1.95) & 12.89 (4.29) & 11.56 (3.29) & 12.12 (2.49) & 3.41 (1.43) &                35.59 (0.94) & 31.4 (1.17) & 10.66 (1.40) & 31.54 (1.56) & 37.25 (1.07) &  6.93 (1.26) & 11.2 (1.74) & 10.66 (1.40) & 10.94 (1.16) & 5.73 (1.11) \\ \hline
R=5 & 7.08 (2.36) & 14.69 (4.47) & 13.38 (3.44) & 7.57 (2.13) & 6.17 (2.05) &                34.00 (1.15) & 30.8 (1.14) & 11.47 (1.42) & 33.67 (1.29) & 34.63 (1.43) &  8.30 (1.35) & 12.0 (1.75) & 11.47 (1.42) & 8.61 (1.24) & 7.74 (1.32) \\ \hline
\end{tabular}}
\caption{\revb{Image space error metrics for the HCP test subjects at different undersampling factors. NMSE values are multipled with $10^3$. }}
\label{tbl:hcp_nmse_psnr_rmse}
\end{table*}

\begin{table*}[t]
\resizebox{\textwidth}{!}{
\begin{tabular}{|l|l|l|l|l|l|l|l|l|l|l|l|l|}
\hline
\multirow{2}{*}{Factor} & \multicolumn{3}{l|}{Absolute k-space error (x$10^3$)} & \multicolumn{3}{l|}{RMSE (\%)} & \multicolumn{3}{l|}{NMSE (x10$^{3}$)} & \multicolumn{3}{l|}{pSNR} \\ \cline{2-13} 
         & l-MALA   & DDP   & VarNet       & l-MALA    & DDP   & VarNet              & l-MALA     & DDP   & VarNet        &      l-MALA     & DDP & VarNet      \\ \hline
R=2 & 64.18 (5.49) &           30.32 (1.82) &           54.64 (1.79) &           10.09 (0.85) &           7.04 (0.61) &           6.70 (0.61) &           10.26 (1.76) &           4.97 (0.87) &           4.52 (0.82) &           35.25 (0.41) &           38.39 (0.46) &           38.82 (0.45) \\ \hline
R=3 & 76.66 (7.56) &           32.54 (2.24) &           64.84 (2.71) &           12.48 (1.38) &           9.60 (1.07) &           9.75 (1.42) &           15.75 (3.46) &           9.33 (2.12) &           9.70 (2.82) &           33.43 (0.45) &           35.70 (0.41) &           35.61 (0.69) \\ \hline
R=4 & 85.42 (9.27) &           33.58 (2.43) &           71.74 (3.50) &           13.80 (1.69) &           11.54 (1.79) &           13.89 (1.84) &           19.34 (4.79) &           13.67 (4.28) &           19.63 (5.22) &           32.56 (0.72) &           34.15 (0.97) &           32.52 (1.04) \\ \hline
R=5 & 93.75 (11.43) &           34.53 (3.25) &           78.31 (4.64) &           15.62 (1.96) &           13.78 (1.86) &           15.38 (2.02) &           24.80 (6.32) &           19.40 (5.47) &           24.11 (6.26) &           31.49 (0.56) &           32.57 (0.81) &           31.63 (0.98) \\ \hline
% R=2 & 64.18 (5.50) &       30.32 (1.82) &       10.11 (0.85) &       7.0 (0.61) &       10.29 (1.76) &       4.98 (0.86) &       35.24 (0.40) &       35.2 (0.46) \\ \hline
% R=3 & 76.66 (7.55) &       32.54 (2.24) &       12.46 (1.39) &       9.6 (1.07) &       15.73 (3.48) &       9.34 (2.12) &       33.44 (0.45) &       33.4 (0.41) \\ \hline
% R=4 & 85.42 (9.29) &       33.58 (2.43) &       13.81 (1.69) &       11.5 (1.78) &       19.36 (4.80) &       13.66 (4.25) &       32.56 (0.72) &       32.6 (0.97) \\ \hline
% R=5 & 93.75 (11.43) &       34.53 (3.25) &       15.62 (1.97) &       13.8 (1.85) &       24.80 (6.35) &       19.42 (5.45) &       31.49 (0.57) &       31.5 (0.82) \\ \hline
\end{tabular}}
\caption{\revb{Mean (std) values for the different metrics for the in-house measured subjects at R=2 to 5. Presented values for abs. error and NMSE are multipled with $10^3$.} }
\label{tbl:metrics_inhouse_multir}
\end{table*}

% \usepackage{multirow}
\begin{table}[]
\begin{tabular}{|l|l|l|l|}
\hline
\multirow{2}{*}{\begin{tabular}[c]{@{}l@{}}Added noise\\ level\end{tabular}} & \multicolumn{3}{l|}{pairwise RMSE (\%)} \\ \cline{2-4} 
                                   & l-MALA       & cWGAN       & local      \\ \hline
x 0                                &  2.32            & 13.28            &  8.02           \\ \hline
x 1                                &  2.42 (0.00)           & 13.26 (0.13)            &  8.02 (0.53)         \\ \hline
x 4                                &  2.43 (0.57)           &  13.53 (0.0)           &  8.04 (0.30)         \\ \hline
x 8                                &  2.51 (0.01)           &   14.62 (0.0)          &  8.00 (0.13)         \\ \hline
\end{tabular}
\caption{Pairwise RMSE for a test subjects from the HCP dataset for changing k-space noise levels at R=5. Number in parenthesis is the p-value of statistical significance of the change in the pairwise RMSE values compared to the previous noise level (using the Wilcoxon signed rank test).} 
\label{tbl:pairwise_rmse_noise}
\end{table}

We start by showing sample images from the latent MALA model in Fig.~\ref{fig:samples} for an image undersampled with factor R=5 alongside the phase and biasfield estimations. The structures in the samples as well as in the mean image, obtained as the mean of the drawn samples, overlap well with the fully sampled image. On the other hand, structural variations between the samples are present, which can also be seen in the std map. Most pixels corresponding to tissue edges have a high std value. This is expected since the missing data in the k-space is mostly in the high-frequency regions, whose contributions are more important for edge pixels. However, it is important to note that the std values on the edges are not homogeneous, indicating some parts of the edges have higher variability. Furthermore, the variations are not limited to edges but also structures in the white matter (WM) as well. In the zoomed region one can see a GM structure inside the WM changing its visibility in the samples (see arrow). This is also captured by the GM segmentations, for which we show the contours. Each contour is from a different segmentation, which itself is from another image sample, from in total 10 samples. The contours show how the network in some cases segments the inlet region as GM and in some cases as WM. The pixel-wise std maps are marginal maps, i.e. they present the variations in the pixels as if they were independent. In reality the variations are not pixel-wise, rather the structures as collections of multiple pixels move between different samples, which can be observed better in the GIFs. \revc{Finally, the MAP image captures a local maximizer of the highly non-convex and multi-modal probability distribution whereas the mean of the samples takes the mean of multiple images from such a distribution, which explains the differences between the two. As such, an artifact arising in the MAP image, for instance, might be averaged out in the mean image.} 
\begin{figure}
    \centering
    \includegraphics[width=0.48\textwidth]{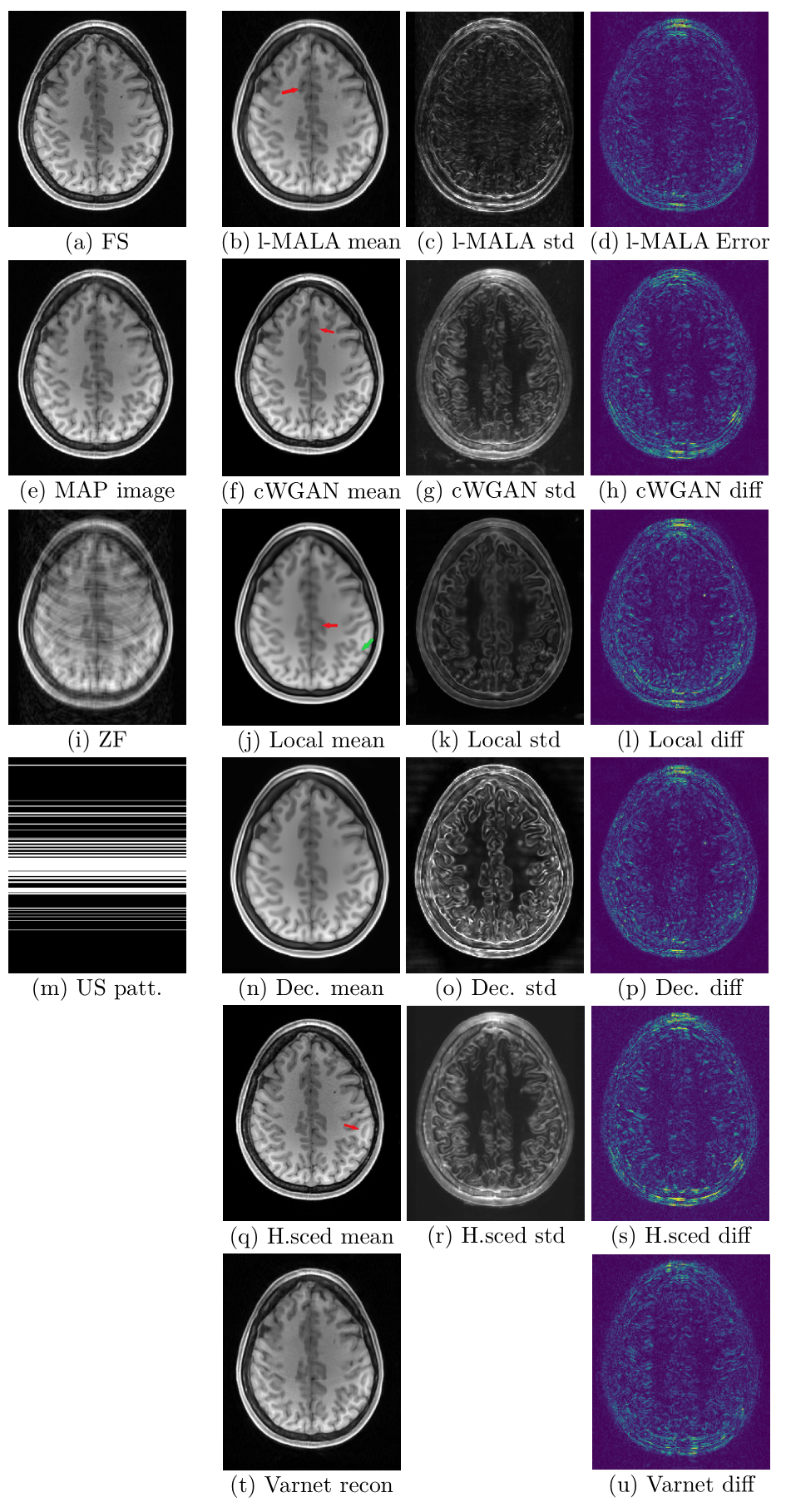}
    \caption{Sampling results for different methods and reconstruction result for the Varnet at R=5. The left most column shows the fully sampled (FS), the MAP estimate, the zero-filled (ZF) images and \revb{the undersampling pattern (US patt.)}. In the rightmost three columns, the sample mean, pixelwise standard deviations and the absolute error maps between the mean and fully sampled images from the respective method are given. \revb{Final row shows the Varnet reconstruction and its error.} The error maps are clipped to (0, 0.3), the std maps are clipped to (0, 0.04) and (0, 0.18) for the l-MALA and the other three sampling methods, respectively.}
    \label{fig:comp}
    \vspace{-0.6cm}
\end{figure}

Fig.~\ref{fig:comp} shows results for the different sampling methods for comparison purposes, namely the latent MALA, the cWGAN, the local VAE sampling \revb{and the reconstruction methods, namely Varnet} and the heterosdecastic network. We use the same undersampling pattern for all the methods for comparability. Both the l-MALA and cWGAN methods capture the underlying image fairly well in the mean of the samples \revb{and the Varnet in the reconstructed image}, which is reflected in the difference images. The mean of the local VAE sampling is very blurry and fails to capture the structures in the underlying image as well as the other methods. The heteroscedastic model performs worse in the mean prediction as expected due to the lack of a data consistency term. The pixelwise standard deviation maps for the VAE and cWGAN models are similar at first glance, in that both reflect the high uncertainty regions at the tissue edges. However, the cWGAN maps are quite noisier and blurrier in comparison. Latent MALA provides a much finer level distinction. This is expected since the proposed method generates samples based on examination of the given data instead of relying on a trained model to generalize and does not make assumptions about data availability from the joint distribution of fully and undersampled images as in cWGAN. The std maps from the heteroscedastic model yields even more blurry results. All methods except the local sampling are capable of indicating some of the regions where their mean maps differ from the ground truth image, by showing high diversity in the samples or high std values in those regions, as exemplified by the arrows on the respective image. The local sampling method also captures variability on the edges. Furthermore, it can also indicate possible differences in its mean and the ground truth images, though only coincidentally. For example, in the region shown with the green arrow, it also assigns high std values, however this is rather due to the fact that there are two edges intersecting heavily in that region. In the region indicated by the red arrow, on the other hand, although the mean map  differs from the ground truth, the local samples do not indicate a high variability in this region. \revb{Finally, comparing the decoder outputs with the proposed sampling method demonstrates the value of taking the sample from $p(x|y,z^t)$ as $x = \mu_{x|y,z^t}$ instead of from $p(x|z^t)$ as $x = \mu_x$. Incorporating the measured k-space data increases the sample quality drastically, increasing data fidelity and showing much more texture than in the decoder output counterpart.}

A critical aspect is the connection between the uncertainty and the direction of the undersampling. As the undersampling is in the posterior-anterior (A-P) direction in contrast to the fully sampled left-right (L-R) readout direction, we expect the uncertainty to be higher in the A-P direction as well. This means, for instance, that the reconstruction should be less certain of the positions of edges lying in the L-R direction and more certain of positions of edges lying in the A-P direction. Looking at the std maps in Figure~\ref{fig:comp} reveals that the proposed l-MALA is the only method that has this directionality. The l-MALA std maps show, for instance, higher values for the posterior and anterior cortical regions and lower values for the left and right parts of the cortex, adhering to the expected directionality. In contrast, the compared methods simply yield similar std values in all directions, not following the directionality of the undersampling.

\revb{Table~\ref{tbl:hcp_abserr_pwrmse} presents the values for quantitative comparison with HCP images. The l-MALA method yields the lowest absolute error in the k-space among sampling methods, i.e. the sum of all absolute differences between the measured k-space and the k-space of the sample, for the measured part of the k-space (see Appendix for the definition). The l-MALA also yields the best results  for the other metrics, however cannot reach the performance of the DDP method, which has a data projection as its final step, achieving zero k-space error and the best NMSE/pSNR values. For completeness we also show image space error metrics in Table~\ref{tbl:hcp_nmse_psnr_rmse}, where l-MALA again performs well.}

\revb{In the lack of ground truth posterior, assessing the quality of samples and uncertainty is a difficult task. To this end, along with the error metrics, we also look at the sample diversity using pairwise RMSE values between the samples in Table~\ref{tbl:hcp_abserr_pwrmse}. Furthermore, we present pairwise RMSE results with increasing k-space noise level in Table~\ref{tbl:pairwise_rmse_noise}. As expected, l-MALA produces more diverse samples when the undersampling ratio and noise levels increase. cWGAN also fulfils this expectation. Local sampling method does not fulfil the expectation for both cases. In the Appendix we also show how the pixel distributions and standard deviation maps as well as the segmentations of }\revb{ the}\revb{ samples change change with changing undersampling ratios and k-space noise levels. These results are also in accordance with the pairwise RMSE experiments and std values of the image samples as well as the diversity of segmentations increase in both cases. }

\begin{figure}[h!]
    \centering
    \includegraphics[width=0.48\textwidth]{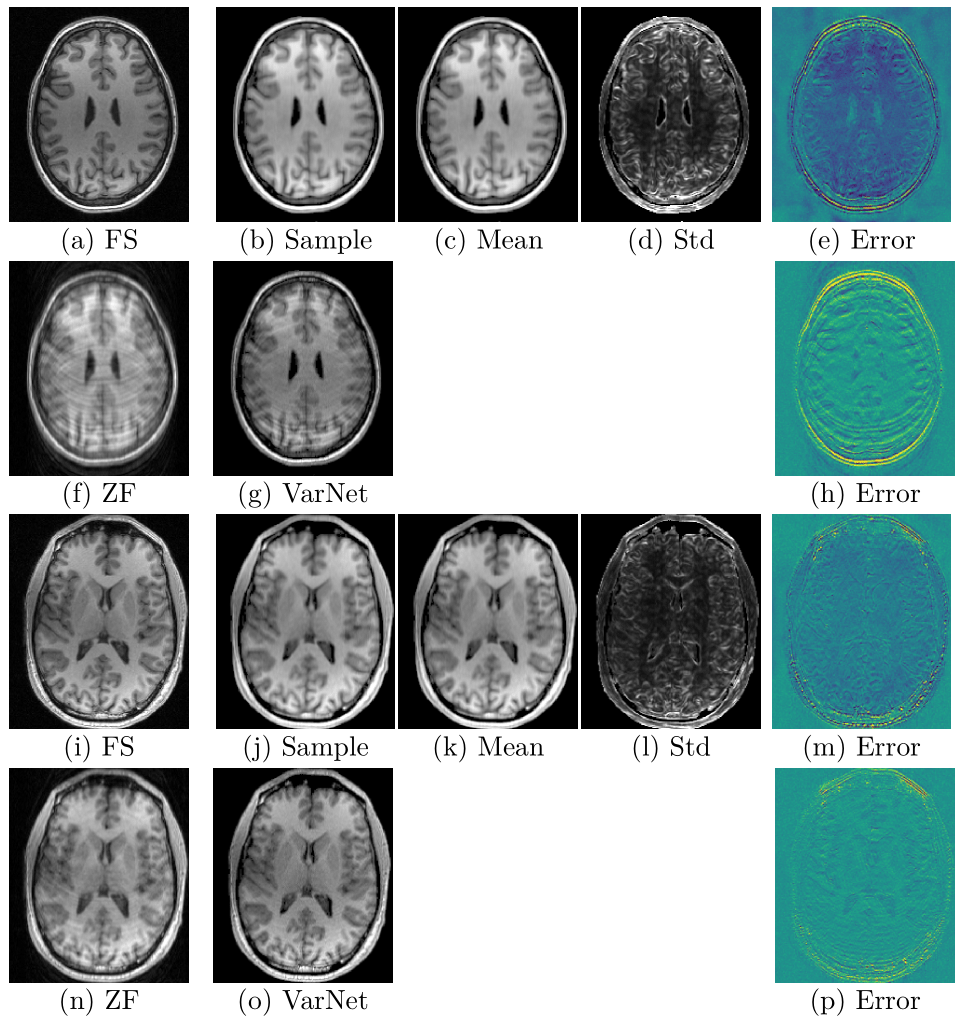}
    \caption{Results for multicoil in-house measured images at R=5 and 3, in the upper and lower blocks, respectively. Shown are the fully sampled (FS), zero-filled (ZF), MAP/Varnet images, the mean and std maps for the latent-MALA samples as well as the difference image between the mean map and the FS (clipped to 0,0.3)). }  
    \label{fig:meas}
\end{figure}

\color{black}

Next, we present results for images from the multicoil in-house measured dataset in Fig.~\ref{fig:meas} for R=3 and 5. We mask out the background in the std map for visual clarity. The method yields similar results for these images as well. The mean map can capture the underlying structures, and most variation is concentrated on the edges, again respecting the directionality of the uncertainty in the phase encoding direction. The smoothly varying error is due to mistakes in the bias field estimation. The VarNet shows remaining artifacts, especially in the first image. We attribute this to the domain shift between the training HCP images and the in-house test images. \revb{We also show quantitative results with the in-house images in Table~\ref{tbl:metrics_inhouse_multir}. The increased error compared to the HCP data is due to multiple factors, such as the domain shift between training and test images, imperfect coil map estimations, and phase estimations... Furthermore, we present more visual results at different undersampling ratios for this data as well in the Appendix.}

\section{Discussion}

\color{black}
The results show that the proposed method is able to capture the underlying ambiguity in undersampled MRI acquisitions in that it generates samples, such as those in Fig.~\ref{fig:samples}, that show realistic structural diversity while retaining high fidelity to the fully sampled image and low k-space error. The variability also indicates potential discrepancies between the mean prediction and the FS image. Further, the proposed method passes the sanity checks with changing noise levels and undersampling ratios. \revb{Again, we emphasize that the aim of the proposed method is not to obtain state-of-the-art reconstruction quality in terms of error metrics, but rather to present a new sampling approach in tackling the inverse problem of MR image reconstruction. }

%Furthermore the variations concentrated in certain regions of the images can be indicative of errors in the mean maps. Hence, along with the mean predictions, the method also provides of indications where it may fail to capture the true tissue structures. Capturing these general variations and especially indications where the mean prediction may fail is very valuable information for any further clinical or research task. This uncertainty information is not provided in the conventional reconstruction approaches.

%Fig.~\ref{fig:samples} shows that the latent MALA method obtains high quality samples, which can represent the underlying image with a high structural fidelity and the method can produce realistic structural variations in the samples. 
\revb{One observation is that the texture in the fully sampled image is not captured in the samples directly from the decoder ($x = \mu_x$) as seen in the upper rows of Fig.~\ref{fig:comp}. This is firstly because we do not add the aleatoric noise on to the samples, of which the texture is partly composed of. Secondly, the lack of texture can be attributed to the VAE, which is known for preferring blurry images and ignore very high-frequency changes. We see that sampling from $p(x|y,z^t)$ instead, improves the sample quality drastically, increasing the texture as well as the structural fidelity, especially at lower undersampling ratios. This improvement is less observable at higher undersampling ratios, as data becomes more sparse at high frequencies in this case. Regardless of data, however, we expect this aspect to improve with a better prior model.} 

\color{black}{}
More importantly, the variation in the samples summarized in the std maps are capable of highlighting the potential mistakes in their mean predictions as seen in Fig.~\ref{fig:comp} for the latent MALA, cWGAN and heteroscedastic feed-forward network approaches. This is valuable information for further decision making, as such regions where uncertainty is high should be approached with doubt.
This information, when taken directly in the form of the samples or estimated standard deviations can be used for any decision making process for clinical or research purposes. The latent MALA and cWGAN are advantageous compared to heteroscedastic models in this respect as they can produce samples allowing quantification of uncertainty of down-stream tasks. 
The heteroscedastic models are limited in that they make a pixel-wise independent Gaussian assumption and cannot generate realistic samples. 
The std maps from the local VAE sampling do not highlight the regions where its mean differs from the ground truth. This is expected, since it takes the MAP estimate as the underlying image and samples only around its latent representation, without exploring other areas of the latent space. Hence it can generate variations only on the tissue edges of the MAP image, but cannot explore possibility of different tissue structures. 

\rest{\textbf{cWGAN}} We note that the cWGAN method was developed for CT and our modified implementation for MR is very straightforward and limited. By feeding only the magnitude images, we set the phase to zero, which is the correct phase for the HCP dataset, giving the method an advantage. However this is not a generic situation and the cWGAN method needs to be extended from CT to work in a realistic MR setting, by incorporating a data consistency term as well as the phase of the image.

\rest{\textbf{Uncertainty and changing noise/undersampling}} As the undersampling ratio R increases the inversion problem becomes harder, hence the mean maps start diverging more from the ground truth.
%Similarly the std maps show higher values with increasing R, as expected. 
Proposed model successfully captures the higher ambiguity for higher undersampling ratios as can be seen in the higher values in the std maps or in the pairwise RMSE values in Table~\ref{tbl:hcp_abserr_pwrmse}. 
This increase is also visible in the increasing spread of the histograms in the Appendix. 
Results presented in Table~\ref{tbl:pairwise_rmse_noise} along with the figures in the Appendix from the experiments with increasing k-space noise $\Sigma_{ns}$, show that the latent MALA model can incorporate the changing k-space noise. When the k-space noise increases ambiguity in the observations increase. Mathematically, the $\Sigma_{ns}$ term in the Gaussian data likelihood increases, which in turn allows accepting samples farther away from the measured k-space data $y$. This results in higher sample diversity, which is reflected in the increase in the standard deviation maps, whereas the mean of the samples is not affected much by this (see also Appendix).

\rest{\textbf{Alternatives to the VAE}} Though we use here the VAE as the latent space model, the outlined method is generic. The integral in Eq.~\ref{eq:marg}, which relates the latent space and the k-space is a generic formulation and can be used with other probabilistic models that provide an explicit likelihood term and a deterministic or probabilistic decoder structure. One necessary property, however, is that the decoder structure needs to be differentiable w.r.t. the latent variables for the Langevin walk. Utilizing another decoder structure here can also increase the quality of the final image samples, for instance models that suffer less from blurriness, such as multi-resolution hierarchical VAE structures~\cite{phiseg}. Furthermore, in case the high dimensions of the image space does not pose problems for sampling, flow based models~\cite{ffjord} can also be used. However, notice that implicit models (e.g. GANs) do not provide an explicit likelihood term, and hence cannot be used in the proposed scheme directly.

\rest{\textbf{Aggregate posterior of the VAE}} Another factor to consider regarding the VAE is that the aggregate posterior of the VAE, given as $q(z) = \int q(z|x) p_{data}(x) dx$ does not necessarily overlap with the prior distribution $p(z)$. This can then cause the random walk to move towards regions of the latent space which are not in the aggregate posterior or similarly miss parts of the aggregate posterior which are zero in the prior. We corrected this discrepancy partially by introducing the empirical prior for $p(z)$ and have not observed problems regarding this issue.

\rest{\textbf{Alternatives to MALA}} Similarly, MALA is not the only way of doing the sampling. We choose it due to several factors, such as its efficiency in high dimensional spaces, theoretical guarantees on asymptotic convergence to the true posterior, not requiring normalized target distributions etc. The target distribution $p(z|y)$ in our formulation is given by a Gaussian, however its covariance matrix is not given in a closed form, rendering direct sampling difficult. Furthermore, the Gaussian posterior is not a generic situation. In cases of more complicated distributions, approaches such as Hamiltonian Monte Carlo~\cite{girolami} can be utilized so that the typical set can be traversed more quickly. Similarly, in cases of multimodal distributions, approaches tailored to such distributions, such as Stein variational gradient descent~\cite{stein_variational} can be considered. Furthermore natural gradient based methods, where the geometry of the target distribution is taken into account by introducing an associated Riemannian metric can be considered to speed up the MALA~\cite{girolami}. One such work is by Pedemento et al.~\cite{pedemento_hmc}, where the authors use Riemannian Hamiltonian Monte Carlo (HMC) to sample from the posterior of emission rates given the photon counts for positron emission tomography. However, they use a uniform, i.e. a non-informative prior for the emission rates, reducing the strength of the model heavily in contrast to using more informative, data-driven priors. Furthermore, the Riemannian metric takes the geometry of the space of probability distributions into account, but not the geometry of the space of the emission rates, as in our case.

\rest{\textbf{MALA and theoretical guarantees}} It is important to emphasize the theoretical guarantees from the choice of MALA for sampling. In contrast to methods such as variational inference, where uncertainty can be underestimated due to the approximate nature of the method~\cite{blei_variational_inference}, MCMC samples are guaranteed to converge to the true distribution, hence to yield the correct statistics. This comes, of course, at the cost of longer inference times. Especially compared to reconstruction methods which can operate orders of magnitude faster. As such, this accuracy/time trade-off has to be evaluated depending on the application. Similarly, the sampling method requires around an hour for MAP estimation and around a second per sample (effectively higher if one considers the auto-correlation of the samples) with our implementation, where only reconstruction can yield an image in sub-second range. Hence, the necessity of sampling at this cost should be evaluated depending on the application.

\rest{\textbf{Modularity}} As discussed above, the proposed method has the advantage of having a modular structure as the prior is decoupled from the data acquisition model and the target posterior is decoupled from the sampling procedure. This, we believe, is quite advantageous in terms of future research and improvement opportunities. This is in contrast to the cWGAN approach, where the loss function and the architecture, by design, determine an implicit target distribution without explicitly modeling the prior or the acquisition. Lastly, the decoupling of the prior from the data acquisition model allows the latent MALA model to be used for different undersampling factors with the same prior without retraining and also incorporate additional details of the acquisition in a very straightforward way, such as the bias field without needing to retrain the prior.

\rest{\textbf{True posterior vs. predictive approaches}} Lastly, the proposed model is sampling from the true posterior for a given $p(x)$ and $y$. Predictive approaches for uncertainty quantification, such as cWGAN, relies on a trained network to generalize for a given sample while having serious training data deficiency. They require training samples that show different fully sampled images for a given undersampled image, which is not readily available.

\revc{\rest{\textbf{Different anatomies/contrasts}}  Another interesting question is how the proposed method would behave for different anatomies and contrasts and other types of images, which we leave for future work.}

\section{Conclusion}
In this paper we proposed and evaluated a method that can provide multiple possible images for the given undersampled k-space data. In contrast to reconstruction approaches, where a single image is output, the sampling approach can capture the uncertainty in the inversion process due to the missing data. The variation in the samples is indicative of the uncertainty, opens up new avenues for uncertainty quantification for following image analysis tasks and can point to potential mistakes in the mean prediction. The method we propose has a modular structure and can be improved by separately improving its components, such as the prior term or the sampling scheme.

\section*{Acknowledgment}
\textcolor{black}{
We thank GyroTools for their MRecon software, NVIDIA for their GPU donation and Klaas P. Pruessmann and Roger C. Luechinger for their help with data acquisition. Thanks to Valery Vishnevskiy and Jonas Adler for sharing their code. }

\textcolor{black}{
\bibliographystyle{IEEEtran}
\bibliography{library.bib}
}

\setcounter{secnumdepth}{4}

\onecolumn
\newpage

\appendix

\title{Appendix to ``Sampling possible reconstructions of undersampled acquisitions in MR imaging with a deep learned prior''}

\author{Kerem C. Tezcan$^\dagger$, Neerav Karani$^\dagger$, Christian F. Baumgartner$^\dagger {}^\ddag$ and Ender Konukoglu$^\dagger$ \\ \\ $^\dagger$ Computer Vision Lab, ETH Zürich, Switzerland \\ $^\ddag$ Medical Image Analysis Group, University of Tübingen, Germany}
\maketitle

\section{Derivation of the closed form of $p(y|z)$}\label{sec:app:first}
As mentioned in the main text the form that is easiest to interpret is given by the marginalization, however this integral difficult to evaluate directly. Instead, we do this by using conjugacy relations for Normal distributions.
We begin by writing 
\begin{equation}
    p(y|x,z)p(x|z) = p(y|z)p(x|y,z).
\end{equation}
Since $p(y|x,z)$ and $p(x|z)$ are Normal distributions, due to the conjugacy, the posterior $p(x|y,z)$ is also a Normal distribution given as $N(\mu_{post},\Sigma_{post})$.
Then\begin{equation}
    p(y|z) = \frac{p(y|x,z)p(x|z)}{N(\mu_{post},\Sigma_{post})}, \hspace{0.5cm} \text{or} \hspace{0.5cm}  p(y|z)N(\mu_{post},\Sigma_{post}) = p(y|x,z)p(x|z).
    \label{eqn:margfrac}
\end{equation} 
Hence the posterior $p(y|z)$ acts as a normalizer to the product distribution to yield a Gaussian. We derive $p(y|z)$ using this relation in Eqn.~\ref{eqn:margfrac}. In the following we also use the conditional independence $p(y|x,z) = p(y|x)$ meaning that when the image is given, this posterior distribution in the k-space is determined without the need for the latent variable. 
For the derivation we use this strategy: i) we first write the product of the two distributions $p(y|x)p(x|z)$, ii) then recognize the mean and covariance of the Normal posterior distribution $N(\mu_{post},\Sigma_{post})$ in this, iii) and separate a Gaussian with these parameters from the whole expression. What is left gives us the target distribution.

The product can be written as
\begin{align}
    & p(y|x)p(x|z)  = \\
    & \det(2\pi \Sigma_{ns})^{-1/2} \det(2\pi \Sigma_{x})^{-1/2} \exp \left \{ -\frac{1}{2}\left[ (y-E x)^H \Sigma_{ns}^{-1} (y-E x) \right] \right \}\\
    & \cdot \exp \left \{ -\frac{1}{2}\left[ (x-\mu_x)^H \Sigma_{x}^{-1} (x-\mu_x) \right] \right \}\\
    & = \det(2\pi \Sigma_{ns})^{-1/2} \det(2\pi \Sigma_{x})^{-1/2} \exp \bigg\{ -\frac{1}{2} x^H\underbrace{(\Sigma_x^{-1} + E^H\Sigma_{ns}^{-1}E)}_\text{$\Sigma_{post}^{-1}$}x \label{eqn:sigmapost} \\ 
    & + Re \{x^H\underbrace{(E^H\Sigma_{ns}^{-1}y + \Sigma_{x}^{-1} \mu_x)}_\text{$\Sigma_{post}^{-1}\mu_{post}$}\} - \frac{1}{2} y^H\Sigma_{ns}^{-1}y - \frac{1}{2} \mu_x^H\Sigma_x^{-1}\mu_x \bigg\} \label{eqn:sigmamupost},
\end{align}
where we have recognized the parameters of the posterior. With these we have enough information to complete the posterior Gaussian. We can replace the terms with posterior parameters and add the missing term $\pm \frac{1}{2} \mu_{post}^H\Sigma_{post}^{-1}\mu_{post}$ to complete the quadratic form as well as the normalizing determinant $\det(2\pi \Sigma_{post})^{\pm 1/2}$, which yields
\begin{align}
    &  = \det(2\pi \Sigma_{ns})^{-1/2} \det(2\pi \Sigma_{x})^{-1/2} \det(2\pi \Sigma_{post})^{+1/2}\det(2\pi \Sigma_{post})^{-1/2}\\
    & \cdot \exp \bigg\{ \underbrace{-\frac{1}{2} x^H\Sigma_{post}^{-1}x + Re\{x^H\Sigma_{post}^{-1}\mu_{post} \}
    - \frac{1}{2} \mu_{post}^H\Sigma_{post}^{-1}\mu_{post} }_\text{$-\frac{1}{2}(x-\mu_{post})^H\Sigma_{post}^{-1}(x-\mu_{post})$}\\ 
    & +  \frac{1}{2} \mu_{post}^H\Sigma_{post}^{-1}\mu_{post}  -\frac{1}{2} y^H\Sigma_{ns}^{-1}y - \frac{1}{2} \mu_x^H\Sigma_x^{-1}\mu_x \bigg\}.
\end{align}
We can combine the quadratic term in the exponent with the determinant term and obtain the complete posterior Gaussian. In this case the expression becomes
\begin{align}
    & p(y|x)p(x|z)  = N(\mu_{post}, \Sigma_{post})\det(2\pi \Sigma_{ns})^{-1/2} \det(2\pi \Sigma_{x})^{-1/2} \det(2\pi \Sigma_{post})^{+1/2}\\
    & \cdot \exp \bigg\{ + \frac{1}{2} \mu_{post}^H\Sigma_{post}^{-1}\mu_{post}  -\frac{1}{2} y^H\Sigma_{ns}^{-1}y - \frac{1}{2} \mu_x^H\Sigma_x^{-1}\mu_x \bigg\}.
\end{align}
Remembering Eqn.~\ref{eqn:margfrac}, we obtain 
\begin{equation}
    p(y|z) = \frac{\det(2\pi \Sigma_{post})^{+1/2}}{\det(2\pi \Sigma_{ns})^{1/2} \det(2\pi \Sigma_{x})^{1/2} }
    \cdot \exp \bigg\{-\frac{1}{2} y^H\Sigma_{ns}^{-1}y  + \frac{1}{2} \mu_{post}^H\Sigma_{post}^{-1}\mu_{post}   - \frac{1}{2} \mu_x^H\Sigma_x^{-1}\mu_x \bigg\}.
\end{equation}

Now taking the logarithm and leaving out the terms that are independent of z we can arrive at the expression we use as 
\begin{equation}
\label{eq:logpost}
    \log p(y|z) =  + \frac{1}{2} \mu_{post}^H\Sigma_{post}^{-1}\mu_{post}   - \frac{1}{2} \mu_x^H\Sigma_x^{-1}\mu_x + C,
\end{equation}
where C denotes some constant with z. Notice that we could leave out the determinant term in the nominator due to our model choice of constant $\Sigma_x$.

Now we need the closed form expression for the first term in the above equation. Also we need to arrive at this using the terms we have access to from the above equations \ref{eqn:sigmapost} and \ref{eqn:sigmamupost}, namely $\Sigma_{post}^{-1}\mu_{post}$ and $\Sigma_{post}^{-1}$. First we write $ \mu_{post} = (\Sigma_{post}^{-1})^{-1}\Sigma_{post}^{-1}\mu_{post}$ and rewrite the target term as $\mu_{post}^H\Sigma_{post}^{-1}\mu_{post} = (\Sigma_{post}^{-1}\mu_{post})^H\mu_{post}$.
Combining the expressions and isoliting the terms constant with z as C then yields
\begin{align}    
     \mu_{post}^H\Sigma_{post}^{-1}\mu_{post} & = \mu_x^H\Sigma_{x}^{-1}(\Sigma_{x}^{-1}+E^H\Sigma_{ns}^{-1}E)^{-1}\Sigma_{x}^{-1}\mu_x \label{eq:post1}\\
     &+2 \text{Re}\left\{ y^H\Sigma_{ns}^{-1}E(\Sigma_{x}^{-1}+E^H\Sigma_{ns}^{-1}E)^{-1}\Sigma_{x}^{-1}\mu_x\right\} + C \label{eq:post2}
\end{align}
Applying the Woodburry identity on the term $(\Sigma_{x}^{-1}+E^H\Sigma_{ns}^{-1}E)$ followed by some algebraic manipulations reveals that this is equivalent to the expression given in~\cite{katarina_marg}.

\section{Derivation of the closed form solution of $p(x|z,y)$ in k-space}
\revb{Here we derive and present the mean and covariance parameters of the posterior distribution $p(x|y,z)$.
First we write
\begin{equation}
    p(x|y,z) = \frac{p(x|z)p(y|x,z)}{p(y|z)} = \frac{p(x|z)p(y|x)}{p(y|z)}, 
\end{equation}
using the model assumption that given the image, k-space is independent of the latent variable, i.e. $p(y|x,z) = p(y|x)$. 
We then write the two distributions on the nominator: i) $p(x|z) = N(x; \mu_x, \Sigma_x)$ and ii) $p(y|x) = N(y; Ex, \Sigma_{ns})$. Since both of these are Normal, the posterior is also Normal due to conjugacy.
However, instead of working in the image space, we prefer to derive the solution in the k-space due to reasons which will be evident later. To this end we write our variable of interest as 
\begin{equation}
    k = FSB\varphi P x,
\end{equation}
with individual terms explained as in the next section. But in essence k is the encoded k-space version of the image variable without the undersampling operator. Notice that as there is no undersampling, we can always recover the image from k as
\begin{equation}\label{eq:k2x}
   x = P^H \varphi^H B^H S^H F^H k, 
\end{equation}
assuming the coil maps are normalized, i.e. $S^HS = I$. Furthermore as the encoding operation is linear the resulting variable k is also Normal distributed.
First we write $p(k|z)$ as 
\begin{equation}
    p(k|z) = N(k; \mu_k, \Sigma_k) =N(k; FSB\varphi P \mu_x, \left[ FSB\varphi P \Sigma_x^{-1}P^H \varphi^H B^H S^H F^H     \right]^{-1}).
\end{equation}
Similarly the data likelihood term becomes
\begin{equation}
    p(y|k) = N(y; Uk, \Sigma_{ns}),
\end{equation}
i.e. y is the noisy observation of the undersampled version of the k-space variable k.
We now write the product again in terms of k as
\begin{equation}
    p(k|z,y) = N(k; \mu_{k|z,y}, \Sigma_{k|z,y}) = p(k|z)p(y|k).
\end{equation}
We can then write the product of these two Normal distributions, complete the square in the exponent and arrive at the resulting mean and covariance matrix as 
\begin{equation}\label{eq:2ndstep_mean}
    \mu_{k|z,y} = \left[ FSB\varphi P \Sigma_x^{-1}P^H \varphi^H B^H S^H F^H + U^H\Sigma_{ns} U \right]^{-1} \left[ FSB\varphi P\Sigma_x^{-1}P^H\varphi^HB^HB\varphi P\mu_x + U^H\Sigma_{ns}^{-1}y \right]
\end{equation}
and
\begin{equation}
    \Sigma_{k|z,y} = \left[ FSB\varphi P \Sigma_x^{-1}P^H \varphi^H B^H S^H F^H + U^H\Sigma_{ns} U \right].
\end{equation}
As also stated in the main text, we implement sampling from this distribution as taking the mean for each latent $z^t$ sample, i.e. $k^t = \mu_{k|z^t,y}$. However, the matrix inversion in Eq.~\ref{eq:2ndstep_mean} is not analytically solvable as the involved matrices are too big, hence we use conjugate gradients to solve the matrix inversion to obtain the solution $\mu_{k|z^t,y}$ for a given $z^t$.  To obtain the image x, we 
then simply take the inverse operations on k as given in Eq.~\ref{eq:k2x} as
\begin{equation}\label{eq:2ndstep_mean_imagespace}
    x^t = P^H \varphi^H B^H S^H F^H k^t.
\end{equation}
To make the expression easier to read, we define the fully sampled encoding operation $E_F \triangleq FSB\varphi P$ by only removing the undersampling operation from the usual encoding operation, i.e. $E = UE_F$. Then we can rewrite Eq.~\ref{eq:2ndstep_mean_imagespace} compactly as 
\begin{equation}
    x^t = E_F^H \left[ E_F\Sigma_x^{-1}E_F^H + U^H\Sigma_{ns}U \right]^{-1} \left[ E_F\Sigma_x^{-1}E_F^H\mu_x + U^H\Sigma_{ns}y \right]
\end{equation}}

\section{Description of the vanilla variational autoencoder (VAE) model}
Here we describe the variational autoencoder model~\cite{KingmaW13, rezendevae} for completeness. 

The VAE is essentially an unsupervised learning based density estimation method. It learns a function called the evidence lower bound (ELBO) that is a lower bound to the target probability density. Here we describe the vanilla VAE and refer the reader to the next section for the description of the 2D latent space architecture.

The basic equation of VAE can be derived by writing
\begin{equation}
    \log p(x) = \log\frac{p(x,z)}{p(z|x)}
\end{equation}
for images $x\in \mathbb{R}^W$ and latent vectors $z\in\mathbb{R}^V$ (generally $V\leq W$) with a simple prior $p(z)$. We can then introduce an auxiliary distribution and rewrite as 
\begin{equation}
    \log p(x) = \log\frac{p(x,z)}{p(z|x)} \frac{q(z|x)}{q(z|x)} = \log\frac{p(x,z)}{q(z|x)}\frac{q(z|x)}{p(z|x)}.
\end{equation}
Then taking an expectation of both sides with $q(z|x)$ yields 
\begin{equation}
    \log p(x) = \mathbb{E}_{q(z|x)}[\log p(x|z)] - KL[q(z|x)||p(z)] + KL[q(z|x)||p(z|x)],
\end{equation}
where KL denotes the Kullback-Leibler divergence. The VAE is trained to maximize the first two terms, called the evidence lower bound (ELBO) with $ELBO(x) = \mathbb{E}_{q(z|x)}[\log p(x|z)] - KL[q(z|x)||p(z)]$, which minimizes the rightmost KL term. After the training, the rightmost KL term becomes small and the ELBO approximates the true distribution, i.e. $\log p(x) \approx ELBO(x)$.

The realization of VAE is done as follows: first an image $x$ is passed through a neural network mapping called the encoder with parameters $\theta^{enc}$ that predicts the distribution $q_{\theta^{enc}}(z|x)$ for the latent variable $z$. This $q(z|x)$ distribution is parameterized using a Normal distribution, i.e. $q(z|x)=N(\mu_{lat}, I\cdot\sigma_{lat})$, where $I$ is the identity matrix, hence the encoder function outputs the two variables $\mu_{lat}\in \mathbb{R}^V$ $\sigma_{lat}\in\mathbb{R}_+^V$.  The decoder mapping is again a neural network with parameters $\theta^{dec}$, which outputs the distribution $p_{\theta^{dec}}(x|z) = N(\mu_{out}, I\cdot \sigma_{out})$. Then the VAE takes samples $z^l \sim q(z|x)=N(\mu_{lat}, I\cdot\sigma_{lat})$ and decodes these using the decoder mapping as $p(x|z^l)$. Then the VAE is trained using training samples to maximize the ELBO by optimizing for the network weights $\theta^{enc}$ and $\theta^{dec}$.

\section{The 2D latent space VAE architecture}
All convolutions are padded and have a kernel size (3, 3) and stride (1, 1) and use a ReLU unless noted otherwise.

The encoder begins with four convolutional layers with 32, 64, 64, 64 output channels, respectively. Then a convolutional layer with kernel size (14, 14), stride (19, 19) and 60 output channels produces the mean of $q(z|x)$ from the fourth layer. Similarly another convolutional layer from the third layer produces the log standard deviation values for $q(z|x)$ with a kernel size of (14, 14), stride (19, 19), without ReLU and 60 output channels. The network is fully convolutional, hence can work with different image sizes. Assuming an input image size of 252x308 for demonstration, the latent space size becomes bx18x22x60, where b is the batch size. We use the usual reparameterization trick to sample $z$'s~\cite{KingmaW13}. At the beginning of the decoder, we apply a scheme of increasing channel dimensions and using these to increase spatial dimensions. We do this in two steps, once for the first image dimension and once again for the second image dimension to obtain a proper reshaping while using the implementation of Tensorflow's reshaping function. First convolutional layer of the decoder does not use ReLU and has 64$\cdot$19=1216 output channels, resulting in a tensor of size bx18x22x1216. The output of this layer is first transposed to bx18x1216x22 and reshaped to bx252x64x22. This layer then gets transposed to bx252x22x64, then goes through a convolutional layer with again 1216 output channels and without ReLU and becomes bx252x22x1216. This then gets reshaped again to yield a tensor size of bx252x308x64, which is the input image size. This tensor then goes through a ReLU. We then apply 6 convolutional layers with each 60 output channels. Finally another convolutional layer with 1 output channel yields the mean prediction.

\section{The extended encoding matrix}\label{sec:encoding}

\rest{As mentioned in the main text, we extend the usual encoding operation in the MR acquisition model to consider additional effects of the image acquisition process.}
Typically, the encoding operation consists of the coil sensitivities~\cite{KlaasP.Pruessmann2001}, the Fourier transform and the undersampling operation.
We include four additional factors in $E$.
% (a) a padding operator, (b) an operation for combining the phase with the magnitude image, (c) an operator for modeling the bias field~\cite{melanie, BiasField_MRI_Sled}, and (d) a scaling factor.
The goal of these extensions is to integrate acquisition-specific knowledge to make the image corresponding to the observed k-space data as similar as possible to the VAE's training images.

Let $\tilde{E} = UFS$ denote the usual MR encoding matrix, where $S:\mathbb{C}^N\rightarrow\mathbb{C}^{Nc}$ is the sensitivity encoding matrix~\cite{KlaasP.Pruessmann2001} with $c$ coils, $F:\mathbb{C}^{Nc} \rightarrow \mathbb{C}^{Nc}$ is the coil-wise Fourier transform and $U: \mathbb{C}^{Nc} \rightarrow \mathbb{C}^{Mc}$ is the undersampling operation.
ven as:
\begin{equation}
E = \tilde{E}B\varphi P s.
\end{equation}
where $P$ is a padding operator, $\varphi$ is an operator that incorporates phase information, $B$ models the bias field in the acquisition and $s$ is a scaling factor. We now describe each of them in more detail.

\subsubsection{\rest{Padding operator, $P$}} The role of $P$ is to minimize any field of view (FOV) differences between the image corresponding to the given k-space data and the space of training images of the VAE. Although our fully convolutional architecture is agnostic to the image size, the empirical prior is estimated for a specific resolution and FOV. Thus, the k-space size can be different due to varying FOV during acquisition. $P$ bridges this gap by padding or cropping the test image to make its size similar to that of the VAE's training images.

\subsubsection{\rest{Phase matrix, $\varphi$}} For computational as well as implementation related purposes, we assume that the phase of structural images is highly independent of the magnitude image and smooth. Hence, the same phase image is used for all our posterior samples. This allows us to separate the magnitude and the phase of the image and run the sampling only on the magnitude of the image. However, note that this assumption is not a requirement for the proposed method (as the phase could be sampled as well) but rather a methodological simplification motivated by empirical observations. Following this assumption, we write the phase as a diagonal matrix $\varphi$ acting on the image.

\subsubsection{\rest{Bias field matrix, $B$}} We use a diagonal matrix, $B$, to explicitly model the bias field in the acquisition~\cite{melanie}. The MR images unavoidably have a bias field due to several factors~\cite{BiasField_MRI_Sled}. However, it is easy to estimate it from the measured data. As the bias varies between different acquisitions, this is a potential source of discrepancy between the test image and the VAE's training images. In order to minimize such a discrepancy, we train the VAE on bias free images. Thus, the samples obtained from the VAE are also free of bias fields. However, as the measured data $y$ has the bias field in it, we estimate this field and apply it to the sampled images. 

\subsubsection{\rest{Scaling factor, $s$}} Finally, we introduce an intensity scale factor to make the data likelihood invariant to any scaling difference between the samples and the k-space. During the random walk in the latent space, the corresponding images might get scaled at each step, meaning the image may be multiplied globally by a scale factor. If this scale factor moves away from 1, this causes the data likelihood to increase, since the scales of the k-space data and the image samples do not match. However, from the perspective of sample quality, this does not pose a problem as long as the scaling factors are known. The sampled images can be brought to the same scale by multiplying them with the inverse of the scaling factor. Furthermore, allowing the scale factor to be different for each sample, allows more freedom to the random walk in the latent space, as it is less constrained by the increase in data likelihood due to scale changes. Hence, such scale invariance is desirable. To this end, we introduce a scalar $s$, that keeps the data likelihood at the lowest, inducing an invariance to scaling. We calculate its value by solving $s^* = \min_s ||Es\mu_x(z^t) - y||_2^2$, where we separate the $s$ term from the extended encoding and use the mean of the decoder as the image. Then, we take the derivative of the expression with respect to $s$ and set it to zero to obtain the minimizing s value, which is given analytically as $s^*=\frac{Re\{\mu_x(z^t)^HE^Hy\}}{\mu_x(z^t)^HE^HE\mu_x(z^t)}$. We do this estimation separately for each $z^t$ sample at each step.

The complex conjugate of the extended encoding operation operation is given as $E^H = s^HP^H\varphi^HB^H\tilde{E}^H$, where we implement $P^H$ as cropping if $P$ is a padding operation and vice versa, $\varphi^H$ is multiplication with the complex conjugate of the phase, $B^H = B$ since the bias field is real and $s^H=s$, again since the scale factor is real.

\section{Measuring the quality of samples}
As we do not have access to the ground truth posterior distribution of images given the k-space data, we resort to using indirect measures and characterize two aspects of the samples. Firstly, the samples have to be in agreement with the measured k-space data. Secondly, the samples have to have a high diversity to the extent allowed by the measured data and the noise in k-space. Notice that there is a trade-off between these two aspects, that is, the measured data constrains the sample diversity and a high sample diversity requires moving away from measured data, increasing the error in k-space.

We use two metrics to quantify the first aspect in Section~\ref{sec:appendix:voxelwise_error}. First is the error in k-space between the samples and the given data for an image. Secondly, though this also considers the parts of the k-space that are not measured, we look at the RMSE between the samples and the original image. We then introduce a pairwise RMSE metric to quantify the sample diversity and present the results in Section~\ref{sec:appendix:sample_diversity}.

\subsection{Distribution of voxelwise error in the measured parts of k-space and NMSE, pSNR and RMSE in the image space}\label{sec:appendix:voxelwise_error}

Here we show the k-space error histograms in Figure~\ref{fig:ksp_err} for a test slice at R=5. We calculate these as follows: we take 50 image samples $\{x_s\}_{s=1}^{50}$ from each method and apply the undersampled Fourier transform to transform each of them to k-space and take the measured voxels. Then we calculate the voxelwise difference between these and the measured data for all measured k-space voxels for all the samples together. The histogram then shows the distribution of the error for all these k-space voxels from all 50 samples. As this difference is complex, we show two histograms seperately for the real and imaginary parts and also for the magnitude values. We can also look at the image-wise k-space absolute error as
\begin{equation}
    \text{absolute error$_s$} = \frac{1}{\text{no of meas. voxels}}\sum_{\text{all meas. voxels}} |Ex_{FS} - Ex_s|,
\end{equation}
for a sample image $x_s$ and the fully sampled image $x_{FS}$. The $|.|$ denotes the magnitude of the complex error value for a pixel and the average is taken over all measured k-space voxels. When calculated for all 50 samples, the mean (std) values for this slice are given as 0.0309 (0.0003), 0.0373 (0.0012) and 0.0381 (0.00047), for the l-MALA, cWGAN and local sampling methods, respectively.

\begin{figure}
    \centering
    \includegraphics[width=0.98\textwidth]{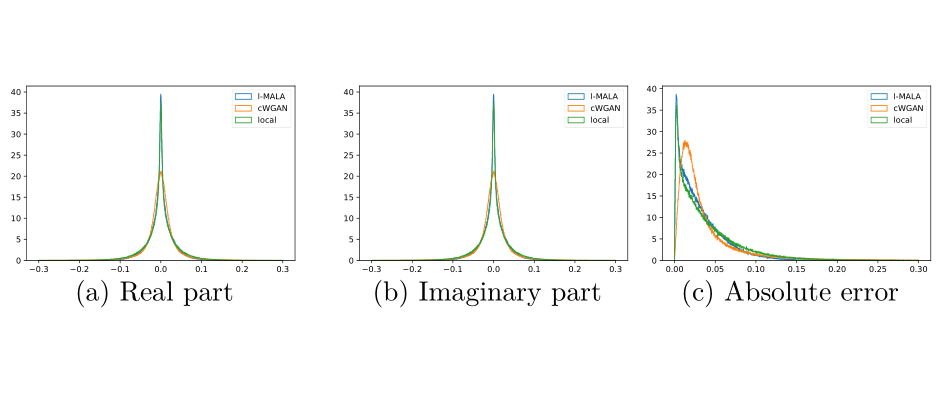}
    \caption{Histograms of the voxelwise error in the measured voxels in the k-space for three different methods for a subject at R=5. As the error is complex, the real and imaginary parts as well as the magnitude of the error values are shown seperately.}
    \label{fig:ksp_err}
\end{figure}

To show how this generalizes, we do a similar analysis using slices from 9 test subjects. We undersample the slices with different patterns for each subject at R=5. Again, for each test subject we generate 50 samples for the three methods each. We then calculate the absolute errors and report the mean and standard deviation values for these in the main text. 
%For all subjects (except subject 3) the l-MALA method yields significantly (p value lower than 0.001) lower absolute error for the 50 samples with the Wilcoxon signed-rank test. For subject 3, the cWGAN method yields significantly lower absolute error (p value lower than 0.001). Considering the mean absolute error for each subject, the l-MALA method yields significantly lower absolute error overall (p value lower than 0.011). The minimum error values follow this trend as well. 
We also calculate the root mean squared error (RMSE) between the 50 samples and the fully sampled image. Though achieving a low RMSE is not the main purpose of any of the methods, we present these results as they still provide some insight into the performance of the methods. To calculate the RMSE we use the formula given in~\cite{tezcan} and use a mask to disregard the background. For in-house images we also perform a bias field correction for both the sample and the fully sampled image before the calculation.
%The RMSE values are significantly lower for the 50 samples for each subject (p value lower than 0.001) and for the mean RMSE values for each subject (p value lower than 0.008) for the l-MALA method compared to the other methods.
Furthermore, again, though sampling is inherently different than only reconstruction, for comparison purposes we present the absolute error and RMSE values of two reconstruction methods, namely the DDP~\cite{tezcan} and the variational network (VarNet)~\cite{hammernikvar}. As the data consistency projection in the DDP method inputs the measured data into the reconstructed k-space, it has a zero absolute error. For both metrics the reconstruction methods yield better performance than sampling methods, which is expected as these are designed to yield the best performance rather than characterize the solution space comprehensively.

\revb{We took the definitions of the normalized mean squared error (NMSE) and  peak signal-to-noise ratio (pSNR) from the fastMRI repository~\cite{fastmri}.}

\subsection{Implementation of the variational network (Varnet) and cWGAN}
For the variational network reconstruction~\cite{hammernikvar} we used the implementation given in \url{https://github.com/visva89/VarNetRecon}. We used a batch size of 2, 48 filters with kernel size 11 at each layer, 10 unfolding layers, filter response as 3.5, 31 knots and cubic interpolation for modeling the activation functions, the $L_2$ loss at the output and otherwise the default parameters. During training we generated undersampled/fully sampled image pairs from the same training set as for the VAE with different patterns at each iteration and fed these into the network. Furthermore the network required an image size of powers of two, for which we padded the images to a size 256x320. We trained for 200000 iterations for R=3,4,5 and 114000 iterations for R=2 with a learning rate of 0.001. At test time we padded the test images (originally 252x308) as well as their undersampling patterns and after reconstruction cropped back to the original size for comparing the performance with different methods.

We trained the cWGAN method~\cite{adler_cwgan} as given in the code shared by the authors. We modified it minimally to work with MR images and trained for 150000/600000/800000/800000 iterations for R=2,3,4,5, respectively, with the decay ratio for the noisy linear cosine decay as 2000000 but otherwise with the default settings in the code provided by the authors and the augmentation used for the VAE.

\begin{table}[]
\resizebox{\textwidth}{!}{
\begin{tabular}{|l|l|l|l|l|l|}
\hline
\multirow{2}{*}{subject}       & \multicolumn{5}{l|}{RMSE (\%)}             \\ \cline{2-6} 
                          & l-MALA & cWGAN & Local & VarNet & DDP \\ \hline
\#1 &  8.5 (8.3, 0.11) & 11.3 (10.7, 0.35) & 11.7 (11.4, 0.13) & 8.7 & 8.2 \\ \hline
\#2 &    9.8 (9.7, 0.06) & 13.4 (12.8, 0.40) & 12.9 (12.6, 0.11) & 9.7 & 9.2 \\ \hline
\#3 &     10.9 (10.9, 0.01) & 15.6 (14.3, 0.76) & 14.5 (14.1, 0.15) & 10.7 & 9.7 \\ \hline
\#4 &     7.8 (7.8, 0.03) & 11.0 (10.4, 0.29) & 10.8 (10.6, 0.10) & 8.2 & 7.4 \\ \hline
\#5 &     7.0 (6.9, 0.03) & 11.3 (10.6, 0.46) & 10.4 (10.1, 0.12) & 8.4 & 6.5 \\ \hline
\#6 &      6.8 (6.8, 0.04) & 9.9 (9.2, 0.35) & 10.2 (10.0, 0.14) & 6.4 & 5.8 \\ \hline
\#7 &      8.0 (8.0, 0.02) & 12.5 (11.3, 0.55) & 11.0 (10.7, 0.12) & 8.3 & 7.7 \\ \hline
\#8 &    9.1 (9.0, 0.07) & 12.8 (12.0, 0.41) & 12.1 (11.9, 0.10) & 9.8 & 9.0 \\ \hline
\#9 &     6.7 (6.6, 0.02) & 10.2 (9.5, 0.42) & 9.7 (9.5, 0.11) & 7.3 & 6.1 \\ \hline
mean (std) &   8.30 (1.35) & 12.0 (1.75) & 11.47 (1.42) & 8.61 (1.24) & 7.74 (1.32) \\ \hline
\end{tabular}}
\caption{RMSE values in percentage for 9 HCP test subjects at R=5. Values shown in format: mean (min, std) for the sampling methods and the single value for the VarNet and DDP reconstruction methods. The last line shows the mean (std) of the 9 subjects.}
\label{tbl:abserror}
\end{table}

% \begin{table}[]
% \begin{tabular}{lll}
%       & pairwise RMSE & pairwise SSIM \\
% l-MALA &               &               \\
% cWGAN  &               &               \\
% local  &               &              
% \end{tabular}
% \caption{Pairwise RMSE and SSIM values showing sample diversity.}
% \end{table}

% \begin{table}[]
% \begin{tabular}{lll}
%             & pairwise RMSE & pairwise SSIM \\
% l-MALA, R=2 &               &               \\
% l-MALA, R=3 &               &               \\
% l-MALA, R=4 &               &               \\
% l-MALA, R=5 &               &              
% \end{tabular}
% \caption{Pairwise RMSE and SSIM values showing sample diversity for different undersampling ratios. }
% \end{table}

% \begin{table}[]
% \begin{tabular}{|l|l|l|}
% \hline
%              & pairwise RMSE & pairwise SSIM \\ \hline
% l-MALA, ns=0 &               &               \\ \hline
% l-MALA, ns=1 &               &               \\ \hline
% l-MALA, ns=4 &               &               \\ \hline
% l-MALA, ns=8 &               &               \\ \hline
% \end{tabular}
% \end{table}

\subsection{Samples and segmentations at different undersampling ratios and k-space noise levels}
\revb{Here we present two figures demonstrating how the changing undersampling ratio and k-space noise levels change the samples. Similarly we present two figures which show how the segmentations change under the same conditions.}

\revb{In Fig.~\ref{fig:usfact}, we show how the statistics from the samples change with changing undersampling ratios. Firstly, we show histograms from three pixels indicated on the FS image for R=2, 3, 4 and 5, from which one can observe that the pixel histograms become wider with increasing R, indicating higher uncertainty. This increase is also reflected in the std maps, which show an increase in std values for increasing R. This result shows that the proposed model is able to capture increasing ambiguity due to higher undersampling ratio.}

\revb{Next we present results in Fig.~\ref{fig:kspnoise} to show the methods sensitivity to the noise in the k-space. The quality of the MAP image degrades due to the high noise. This is reflected less in the mean maps, however the standard deviation values increase. This is how the model should behave since the added noise increases the values in $\Sigma_{ns}$, which then allows samples to move farther away from the measured data and show higher diversity. This is also reflected in the histograms of three pixel's intensities, which are indicated in the top std map, as the distributions become wider with increasing noise.} 

\revb{In Figures~\ref{fig:segm_kspnoise} and \ref{fig:segm_usfact} we show the segmentation results for the same settings. In the final row of each figure, we show the pixels where the binarized standard deviation map, i.e. 1 if there is variation in that pixel among samples, 0 if there is no variation in that pixel among samples. We do this for visualisation purposes and as the segmentation maps are binary, their standard deviation values are not informative in any case. These also confirm the observations from Figures~\ref{fig:kspnoise} and \ref{fig:usfact}, that the samples behave as expected from the theory.}

\begin{figure}[th!]
    \centering
    \includegraphics[width=0.98\textwidth]{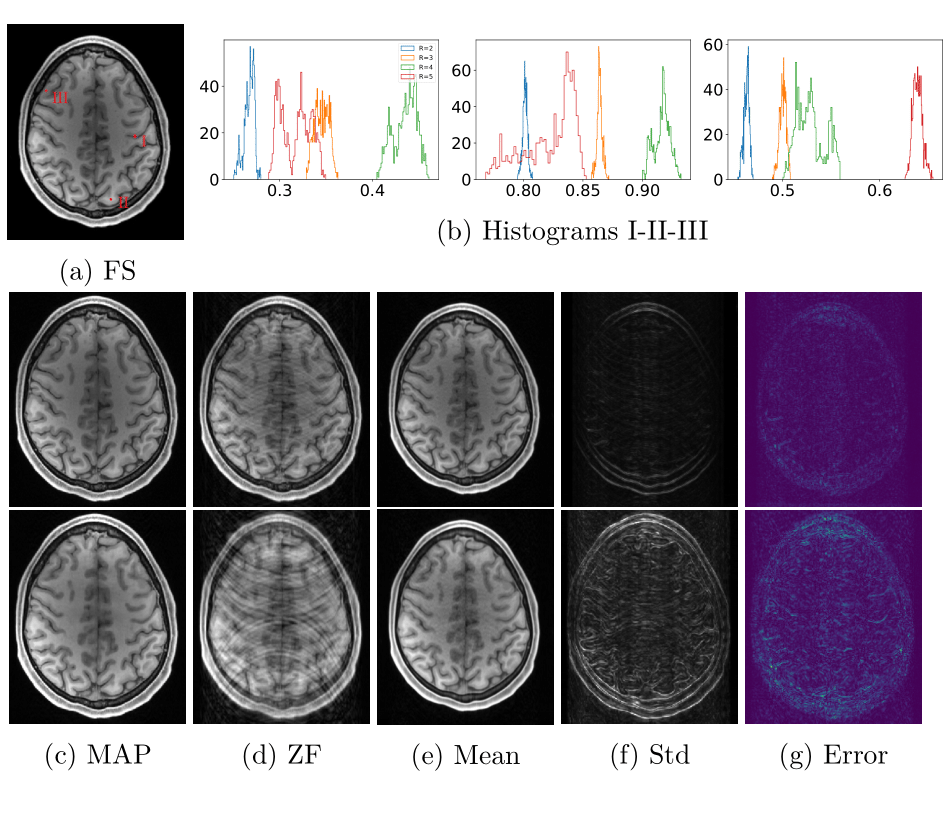} 
    \caption{Results for changing undersampling ratios. First row shows the fully sampled image (FS) and histograms of pixels values in all samples for the pixels indicated on the FS image as I, II and III, respectively. Note the different bin positions for the histograms. Rows two and three show results for R=2 and R=4, respectively. Each row shows the MAP estimation, the zero filled image (ZF), the pixelwise mean and standard deviation maps and the absolute error map between the mean and the FS image (clipped to (0,0.3)).}
    \label{fig:usfact}
    \vspace{-0.6cm}
\end{figure}

\begin{figure}[th!]
    \centering
    \includegraphics[width=0.98\textwidth]{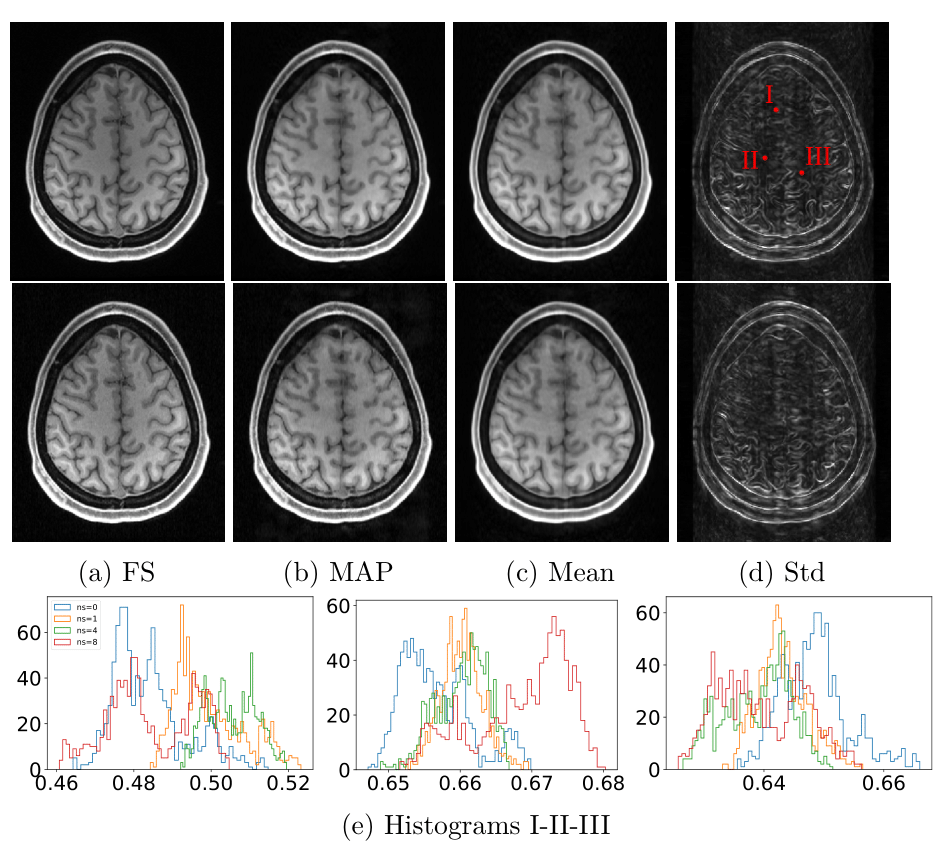}
    \caption{Results for changing the noise in k-space at R=5. First row shows the results with the basis HCP k-space noise. Second row shows the results with noise added on the k-space with 8 times the original noise standard deviation. Third row shows histograms of values of the pixels indicated on the std map (with added noise 1, 4 and 8 times of the basis noise). Note that the fully sampled (FS) image also changes due to the added noise. }
     \label{fig:kspnoise}
\end{figure}

\begin{figure}[th!]
    \centering
    \includegraphics[width=0.98\textwidth]{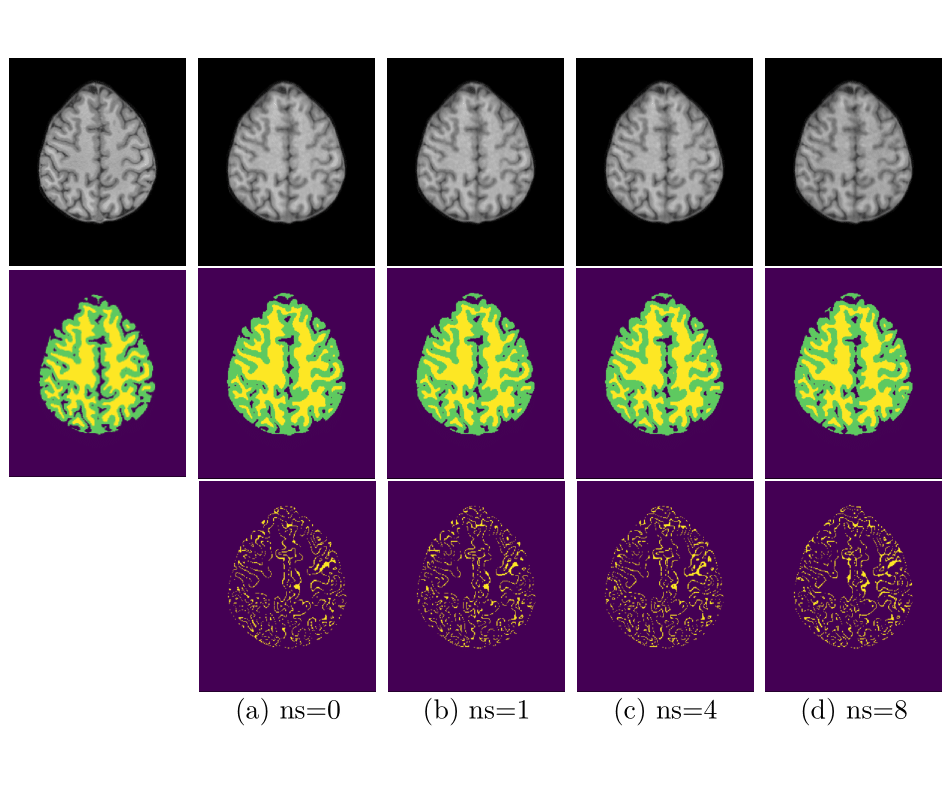}
    \caption{\revb{Segmentation results for changing the noise in k-space at R=5. Leftmost column show the fully-sampled image and its segmentation. First row shows a random sample at increasing added k-space noise levels. Second row shows the segmentation of the corresponding sample in the above row. Last row shows the pixels where there is variation in the segmentations for all samples.}}
     \label{fig:segm_kspnoise}
\end{figure}

\begin{figure}[th!]
    \centering
    \includegraphics[width=0.98\textwidth]{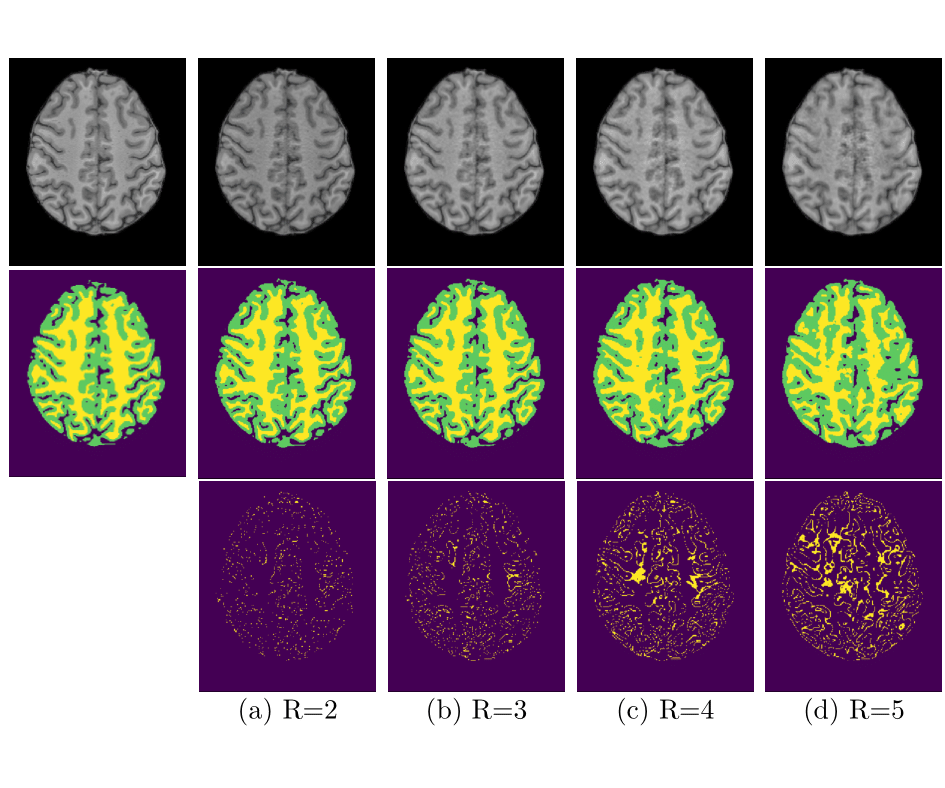}
    \caption{\revb{Segmentation results for changing the undersampling ratios. Leftmost column show the fully-sampled image and its segmentation. First row shows a random sample at varying undersampling ratios. Second row shows the segmentation of the corresponding sample in the above row. Last row shows the pixels where there is variation in the segmentations for all samples.}}
     \label{fig:segm_usfact}
\end{figure}

\subsection{An in-house measured image at different undersampling ratios}
\revb{Here we present an image at different undersampling ratios in Figure~\ref{fig:meas_multi}.}

\begin{figure}[h!]
    \centering
    \includegraphics[width=0.98\textwidth]{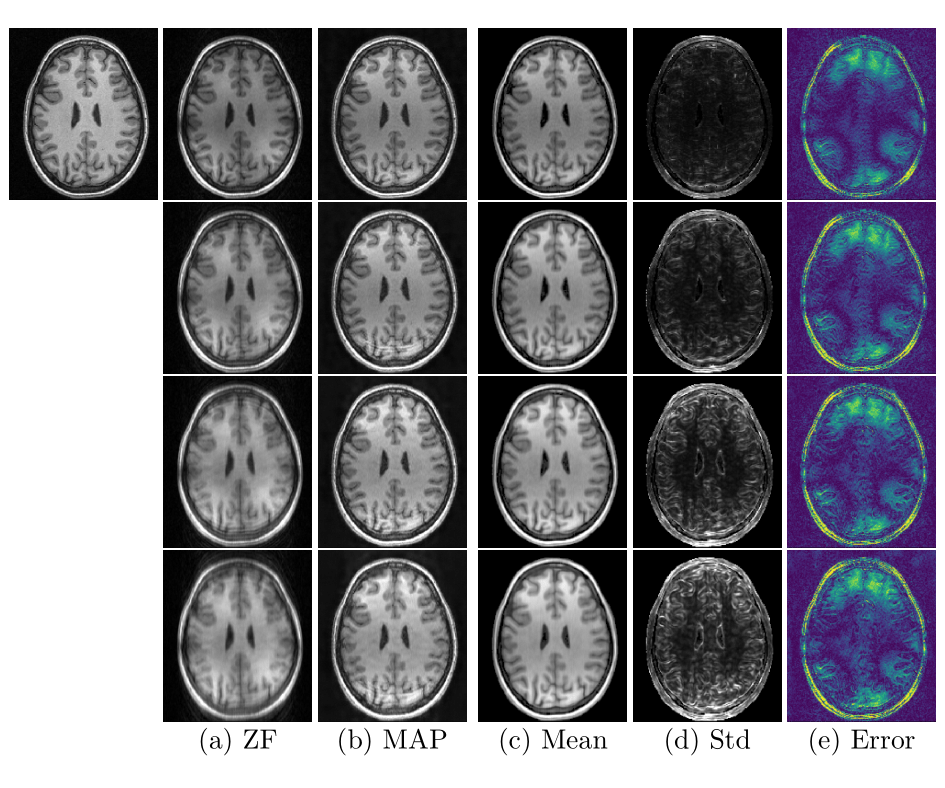}
    \caption{\revb{Sampling results for the in-house measured images for changing undersampling ratios at R=2,3,4 and 5 from top down, respectively.}}
     \label{fig:meas_multi}
\end{figure}

\subsection{Comparison of sample diversity using pairwise RMSE at different noise rates and undersampling ratios}\label{sec:appendix:sample_diversity}

In this section we introduce the pairwise RMSE metric, which we use to measure the sample diversity at different noise levels and undersampling ratios for different methods~\cite{anna_empprior} (We use RMSE instead of the structural similarity index measure as in the reference as the first reflects structural changes better). For this metric we take 1000 pairs of random samples from a method at a noise level or undersampling ratio for a subject and calculate the RMSE between these pairs. This yields 1000 RMSE values, of which we then take the mean to obtain the pairwise RMSE value for this subject and for the method at this noise level or undersampling ratio. 

The aim here, again, to verify the behavior of samples with changing setting, i.e. the "sanity check" experiment. The main idea is that if the k-space noise level is higher, this means that images that are possible solutions to the inverse problem can be farther away from the measured k-space data, which allows these images to be more different than each other, i.e. more diverse. Similarly, if the undersampling ratio is higher, there is less measured data that determine the solutions to the inverse problem, which again allows solution images to be more different from each other leading to higher diversity. As this is a basic relationship between the measurement setting and diversity of solutions, any sampling algorithm should also adhere to this relationship and hence we use this as a "sanity check". 

To this end we added noise on the k-space of an HCP image at R=5 similar to the experiment in the main text. We also experimented with varying the undersampling ratio similar the experiment in the main text at the base noise level. We show these in the main text. One can see that for both l-MALA and cWGAN the pairwise RMSE, i.e. sample diversity is increasing with increasing noise levels in the k-space, as expected. The same trend is not observed for the local sampling method, meaning that the method does not fulfil the expectation. A similar conclusion can be made for the results of changing the undersampling ratio as seen in the main text.

\section{Error metrics for the in-house measured images}
\revb{In Tables~\ref{tbl:metrics_inhouse_r2}, \ref{tbl:metrics_inhouse_r3}, \ref{tbl:metrics_inhouse_r4} and ~\ref{tbl:metrics_inhouse_r5} we present the used metrics for 6 subjects from the in-house measured dataset at R=2 to 5, respectively.} The absolute error is shown as the average of all coils. The DDP absolute error is not zero as the final inverse-forward encoding operations change the k-space in case of multiple coils. Multiple factors contribute to higher error values compared to the HCP images, such as the domain shift between the HCP training images and in-house test images, errors in coil sensitivity estimations from ESPIRiT, or higher errors in the used phase from the MAP estimate. Another observation is that the RMSE values for the l-MALA samples are higher than for the DDP reconstructions. This is mostly because the DDP has a data projection inputting the measured data to the reconstructed k-space, whereas sampling allows for some distance to the measured data to account for the noise.

% \usepackage{multirow}
\begin{table}[]
\resizebox{\textwidth}{!}{%
\begin{tabular}{|l|l|l|l|l|l|l|l|l|l|l|l|l|}
\hline
\multirow{2}{*}{subject} & \multicolumn{3}{l|}{absolute k-space error (x10$^3$)} & \multicolumn{3}{l|}{RMSE ($\%$)} & \multicolumn{3}{l|}{NMSE (x10$^{3}$)} & \multicolumn{3}{l|}{pSNR} \\ \cline{2-13} 
               & l-MALA       & DDP      & VarNet                     & l-MALA     & DDP   & VarNet             & l-MALA      & DDP    & VarNet    & l-MALA           & DDP   & VarNet     \\ \hline
\#1 &   61.16 (61.09, 0.05)  & 29.24  & 55.34 &                                    9.47 (9.40, 0.04)  & 6.20  & 6.01 &                                    8.98 (8.84, 0.08)  & 3.85  & 3.61  &                                    34.81 (34.87, 0.04)  & 38.48  & 38.76                   \\ \hline
\#2 &   60.14 (60.06, 0.06)  & 28.21  & 52.60 &                                    9.83 (9.63, 0.13)  & 6.78  & 7.16 &                                    9.66 (9.28, 0.25)  & 4.60  & 5.12  &                                    35.48 (35.65, 0.11)  & 38.70  & 38.24                   \\ \hline
\#3 &   68.13 (68.07, 0.03)  & 33.85  & 56.45 &                                    9.06 (8.99, 0.03)  & 6.84  & 6.10 &                                    8.21 (8.08, 0.05)  & 4.67  & 3.71  &                                    35.00 (35.07, 0.03)  & 37.45  & 38.44                   \\ \hline
\#4 &   67.46 (67.40, 0.03)  & 31.08  & 54.79 &                                    11.49 (11.44, 0.03)  & 8.21  & 7.61 &                                    13.19 (13.08, 0.07)  & 6.66  & 5.78  &                                    35.97 (36.01, 0.02)  & 38.94  & 39.56                   \\ \hline
\#5 &   72.14 (72.09, 0.03)  & 30.37  & 56.70 &                                    10.96 (10.90, 0.06)  & 7.29  & 7.07 &                                    12.01 (11.88, 0.14)  & 5.30  & 4.98  &                                    34.89 (34.94, 0.05)  & 38.45  & 38.71                   \\ \hline
\#6 &   56.06 (56.01, 0.02)  & 29.20  & 51.98 &                                    9.76 (9.70, 0.03)  & 6.89  & 6.25 &                                    9.52 (9.41, 0.06)  & 4.75  & 3.90  &                                    35.33 (35.38, 0.03)  & 38.35  & 39.21                   \\ \hline
mean (std) & 64.18 (5.49) &           30.32 (1.82) &           54.64 (1.79) &           10.09 (0.85) &           7.04 (0.61) &           6.70 (0.61) &           10.26 (1.76) &           4.97 (0.87) &           4.52 (0.82) &           35.25 (0.41) &           38.39 (0.46) &           38.82 (0.45) \\ \hline
\end{tabular}%
}
\caption{\revb{Different metrics for the 6 in-house measured subjects at R=2. } }
\label{tbl:metrics_inhouse_r2}
\end{table}

% \usepackage{multirow}
\begin{table}[]
\resizebox{\textwidth}{!}{
\begin{tabular}{|l|l|l|l|l|l|l|l|l|l|l|l|l|}
\hline
\multirow{2}{*}{subject} & \multicolumn{3}{l|}{absolute k-space error (x10\textasciicircum{}3)} & \multicolumn{3}{l|}{RMSE (\%)} & \multicolumn{3}{l|}{NMSE (x10$^{3}$)} & \multicolumn{3}{l|}{pSNR} \\ \cline{2-13} 
               & l-MALA       & DDP      & VarNet                     & l-MALA     & DDP   & VarNet             & l-MALA      & DDP    & VarNet    & l-MALA           & DDP   & VarNet     \\ \hline
\#1 &   72.66 (72.52, 0.14)  & 30.78  & 65.95 &                                    11.04 (10.82, 0.12)  & 8.35  & 8.14 &                                    12.20 (11.72, 0.26)  & 6.97  & 6.61  &                                    33.48 (33.65, 0.09)  & 35.91  & 36.13                   \\ \hline
\#2 &   72.20 (72.09, 0.09)  & 30.58  & 62.63 &                                    13.50 (13.40, 0.15)  & 10.25  & 11.28 &                                    18.23 (17.95, 0.41)  & 10.53  & 12.72  &                                    32.72 (32.79, 0.09)  & 35.10  & 34.28                   \\ \hline
\#3 &   80.91 (80.72, 0.12)  & 35.46  & 65.59 &                                    10.92 (10.69, 0.15)  & 8.88  & 8.55 &                                    11.92 (11.42, 0.32)  & 7.88  & 7.30  &                                    33.38 (33.57, 0.12)  & 35.18  & 35.51                   \\ \hline
\#4 &   84.91 (84.76, 0.07)  & 35.76  & 68.25 &                                    14.42 (14.14, 0.14)  & 11.54  & 11.73 &                                    20.81 (20.01, 0.41)  & 13.29  & 13.77  &                                    33.99 (34.16, 0.09)  & 35.94  & 35.78                   \\ \hline
\#5 &   84.96 (84.74, 0.12)  & 32.15  & 66.58 &                                    13.47 (13.38, 0.07)  & 9.73  & 10.26 &                                    18.15 (17.90, 0.20)  & 9.45  & 10.52  &                                    33.09 (33.15, 0.05)  & 35.93  & 35.46                   \\ \hline
\#6 &   64.34 (64.26, 0.05)  & 30.52  & 60.06 &                                    11.49 (11.37, 0.06)  & 8.87  & 8.53 &                                    13.21 (12.93, 0.13)  & 7.85  & 7.27  &                                    33.91 (34.00, 0.04)  & 36.17  & 36.50                   \\ \hline
mean (std) & 76.66 (7.56) &           32.54 (2.24) &           64.84 (2.71) &           12.48 (1.38) &           9.60 (1.07) &           9.75 (1.42) &           15.75 (3.46) &           9.33 (2.12) &           9.70 (2.82) &           33.43 (0.45) &           35.70 (0.41) &           35.61 (0.69) \\ \hline
\end{tabular}}
\caption{\revb{Different metrics for the 6 in-house measured subjects at R=3. } }
\label{tbl:metrics_inhouse_r3}
\end{table}

% \usepackage{multirow}
\begin{table}[]
\resizebox{\textwidth}{!}{
\begin{tabular}{|l|l|l|l|l|l|l|l|l|l|l|l|l|}
\hline
\multirow{2}{*}{subject} & \multicolumn{3}{l|}{absolute k-space error (x10\textasciicircum{}3)} & \multicolumn{3}{l|}{RMSE (\%)} & \multicolumn{3}{l|}{NMSE (x10$^{3}$)} & \multicolumn{3}{l|}{pSNR} \\ \cline{2-13} 
               & l-MALA       & DDP      & VarNet                     & l-MALA     & DDP   & VarNet             & l-MALA      & DDP    & VarNet    & l-MALA           & DDP   & VarNet     \\ \hline
\#1 &   80.38 (80.21, 0.13)  & 31.06  & 72.78 &                                    11.72 (11.54, 0.13)  & 9.02  & 11.38 &                                    13.73 (13.31, 0.30)  & 8.13  & 12.95  &                                    32.96 (33.10, 0.09)  & 35.24  & 33.22                   \\ \hline
\#2 &   82.53 (82.43, 0.07)  & 33.03  & 71.21 &                                    14.02 (13.79, 0.18)  & 11.35  & 14.80 &                                    19.66 (19.02, 0.50)  & 12.92  & 21.91  &                                    32.39 (32.53, 0.11)  & 34.22  & 31.93                   \\ \hline
\#3 &   87.19 (87.06, 0.13)  & 36.20  & 69.90 &                                    12.52 (12.36, 0.08)  & 10.92  & 13.42 &                                    15.68 (15.28, 0.20)  & 11.93  & 18.02  &                                    32.19 (32.30, 0.06)  & 33.37  & 31.58                   \\ \hline
\#4 &   98.13 (97.94, 0.10)  & 37.39  & 77.40 &                                    15.26 (15.04, 0.11)  & 13.02  & 14.54 &                                    23.28 (22.63, 0.33)  & 16.96  & 21.13  &                                    33.50 (33.62, 0.06)  & 34.88  & 33.92                   \\ \hline
\#5 &   94.31 (94.13, 0.16)  & 32.86  & 73.28 &                                    16.58 (16.40, 0.12)  & 14.53  & 17.00 &                                    27.50 (26.91, 0.41)  & 21.21  & 28.91  &                                    31.29 (31.38, 0.06)  & 32.42  & 31.07                   \\ \hline
\#6 &   69.98 (69.87, 0.04)  & 30.94  & 65.90 &                                    12.72 (12.58, 0.08)  & 10.42  & 12.19 &                                    16.18 (15.81, 0.19)  & 10.87  & 14.86  &                                    33.03 (33.13, 0.05)  & 34.76  & 33.40                   \\ \hline
mean (std) & 85.42 (9.27) &           33.58 (2.43) &           71.74 (3.50) &           13.80 (1.69) &           11.54 (1.79) &           13.89 (1.84) &           19.34 (4.79) &           13.67 (4.28) &           19.63 (5.22) &           32.56 (0.72) &           34.15 (0.97) &           32.52 (1.04) \\ \hline
\end{tabular}}
\caption{\revb{Different metrics for the 6 in-house measured subjects at R=4. } }
\label{tbl:metrics_inhouse_r4}
\end{table}

% \usepackage{multirow}
\begin{table}[]
\resizebox{\textwidth}{!}{
\begin{tabular}{|l|l|l|l|l|l|l|l|l|l|l|l|l|}
\hline
\multirow{2}{*}{subject} & \multicolumn{3}{l|}{absolute k-space error (x10\textasciicircum{}3)} & \multicolumn{3}{l|}{RMSE (\%)} & \multicolumn{3}{l|}{NMSE (x10$^{3}$)} & \multicolumn{3}{l|}{pSNR} \\ \cline{2-13} 
               & l-MALA       & DDP      & VarNet                     & l-MALA     & DDP   & VarNet             & l-MALA      & DDP    & VarNet    & l-MALA           & DDP   & VarNet     \\ \hline
\#1 &   85.07 (84.97, 0.07)  & 31.60  & 78.71 &                                    14.91 (14.72, 0.18)  & 14.33  & 16.46 &                                    22.24 (21.67, 0.54)  & 20.63  & 27.15  &                                    30.87 (30.98, 0.10)  & 31.19  & 30.00                   \\ \hline
\#2 &   89.32 (89.20, 0.10)  & 34.01  & 77.06 &                                    16.42 (16.16, 0.15)  & 13.99  & 16.05 &                                    26.97 (26.11, 0.49)  & 19.67  & 25.79  &                                    31.02 (31.16, 0.08)  & 32.39  & 31.22                   \\ \hline
\#3 &   97.16 (97.05, 0.05)  & 36.41  & 76.11 &                                    13.59 (13.41, 0.08)  & 12.11  & 13.60 &                                    18.47 (17.99, 0.22)  & 14.69  & 18.50  &                                    31.48 (31.60, 0.05)  & 32.48  & 31.47                   \\ \hline
\#4 &   109.11 (109.04, 0.03)  & 40.62  & 86.56 &                                    19.08 (18.85, 0.16)  & 17.36  & 18.60 &                                    36.40 (35.52, 0.60)  & 30.32  & 34.70  &                                    31.56 (31.66, 0.07)  & 32.35  & 31.76                   \\ \hline
\#5 &   105.48 (105.18, 0.22)  & 33.63  & 80.25 &                                    16.38 (16.03, 0.25)  & 13.19  & 15.26 &                                    26.84 (25.69, 0.83)  & 17.47  & 23.31  &                                    31.40 (31.59, 0.13)  & 33.26  & 32.01                   \\ \hline
\#6 &   76.34 (76.18, 0.09)  & 30.92  & 71.17 &                                    13.36 (13.28, 0.06)  & 11.67  & 12.33 &                                    17.86 (17.63, 0.17)  & 13.65  & 15.21  &                                    32.60 (32.66, 0.04)  & 33.77  & 33.30                   \\ \hline
mean (std) & 93.75 (11.43) &           34.53 (3.25) &           78.31 (4.64) &           15.62 (1.96) &           13.78 (1.86) &           15.38 (2.02) &           24.80 (6.32) &           19.40 (5.47) &           24.11 (6.26) &           31.49 (0.56) &           32.57 (0.81) &           31.63 (0.98) \\ \hline
\end{tabular}}
\caption{\revb{Different metrics for the 6 in-house measured subjects at R=5. } }
\label{tbl:metrics_inhouse_r5}
\end{table}

\section{Convergence of the chain}
Here we take a random HCP image and show the mean intensity in the brain in this image (i.e. after masking) throughout the MCMC iterations. The plot in Figure~\ref{fig:mcmc_signal_intensity} shows that the chain converges to a range of values close to its initialization. A long burn-in period seems to be avoided by initializing with the MAP estimate. Theoretically, with infinitely many steps the MCMC chain has to discover all solutions. It is, however, possible that for finite chains the initialization introduces some bias. Though it is difficult to show here that this is not the case, the intensities seem to vary throughout the chain, providing empirical evidence that the chain can explore freely.

\begin{figure}
    \centering
    \includegraphics[width=0.48\textwidth]{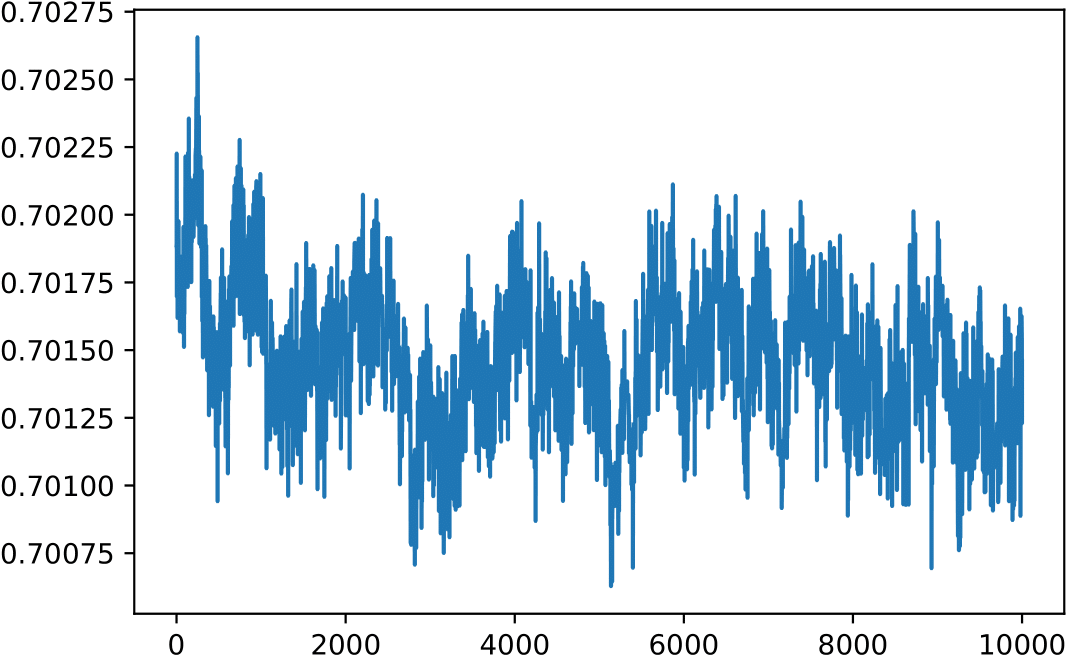}
    \caption{Change of mean signal intensity in the brain throughout MCMC iterations for a random HCP image showing convergence of the chain.}
    \label{fig:mcmc_signal_intensity}
\end{figure}

\section{Comparing the samples $x^t \sim p(x|z^t)$ vs $x^t \sim p(x|y, z^t)$ }
Here we compare the sample quality for the proposed l-MALA with the samples as direct outputs of the decoder. As seen in Figure~\ref{fig:2ndstep_vs_dec}, the proposed sampling approach improves the sample quality drastically.
\begin{figure}[h!]
    \centering
    \includegraphics[width=0.98\textwidth]{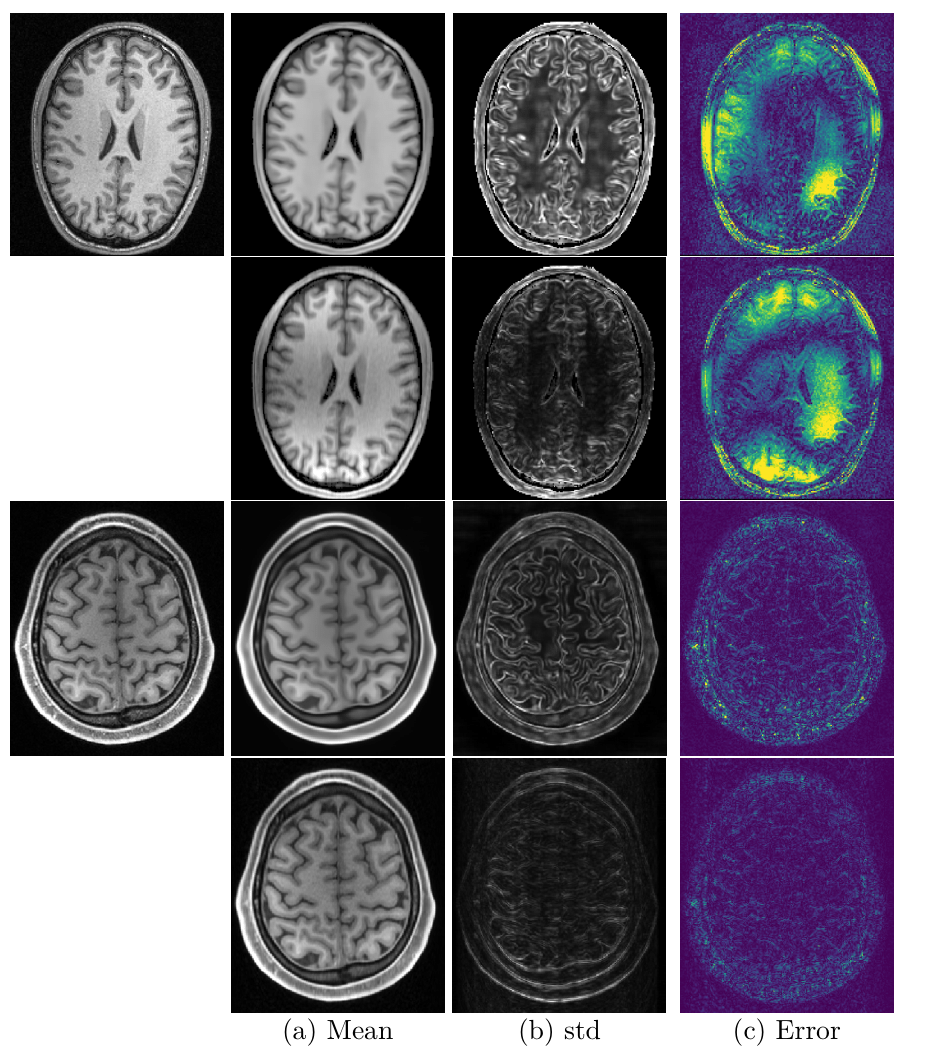}
    \caption{\revb{Comparison between the proposed sampling versus the simpler alternative of taking the decoder output at R=3. The upper block shows an in-house measured image, the lower block shows an HCP image. The leftmost column presents the full-sampled images. In each block the upper row represents the decoder output ($x \sim p(x|z^t)$) and the lower row represents the sample from $x\sim p(x|y,z^t)$.}}  
    \label{fig:2ndstep_vs_dec}
    \vspace{-0.6cm}
\end{figure}

\section{More comparisons on the HCP data}
Here we provide more figures for comparing with the alternative methods for different images and undersampling ratios in Figures~\ref{fig:morecomp1}-\ref{fig:morecomp7}.

\begin{figure}[h!]
    \centering
    \includegraphics[width=0.77\textwidth]{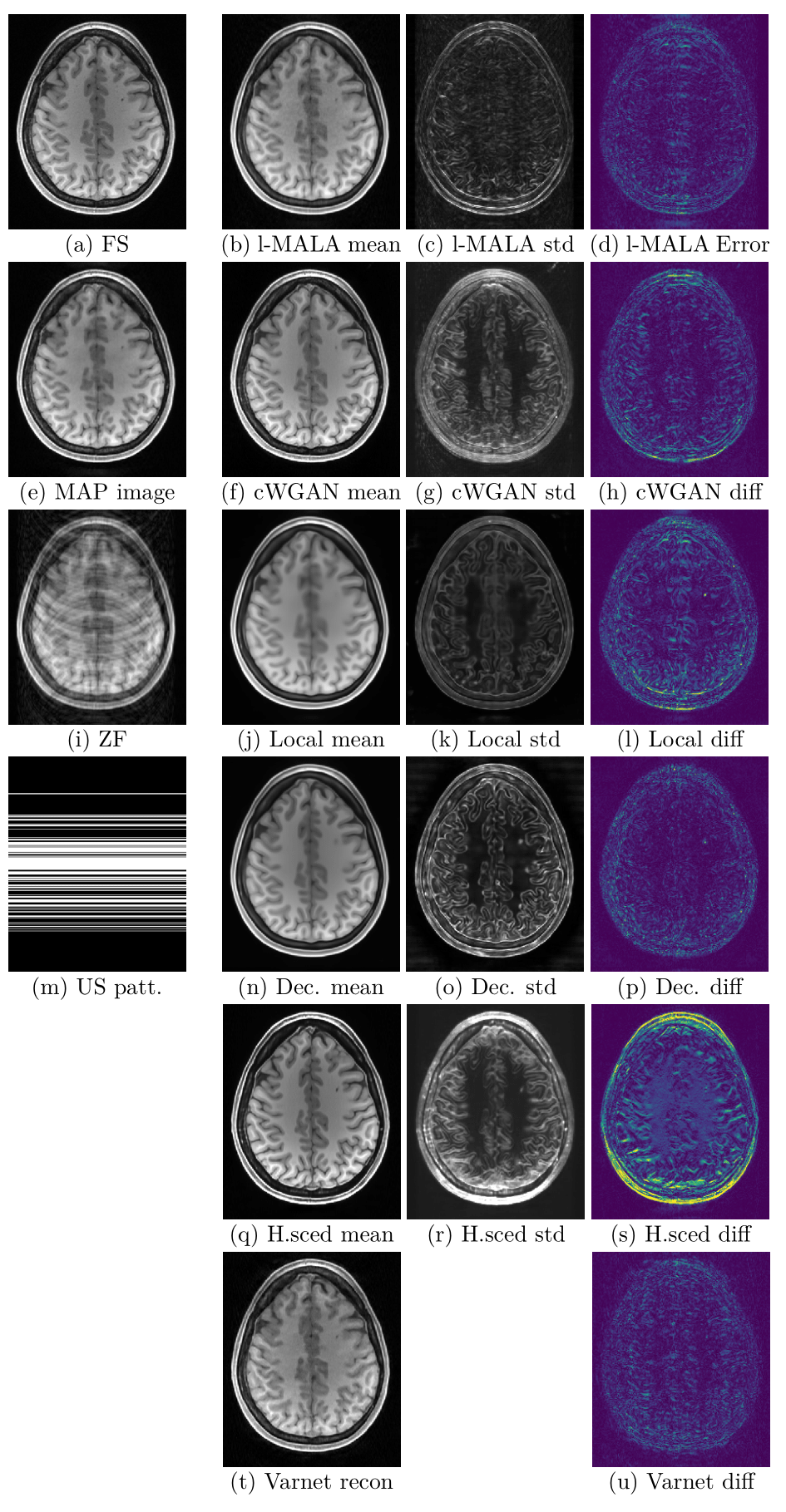}
    \caption{Comparisons at R=4. Figure description same as in main text.}  
    \label{fig:morecomp1}
    \vspace{-0.6cm}
\end{figure}

\begin{figure}[h!]
    \centering
    \includegraphics[width=0.77\textwidth]{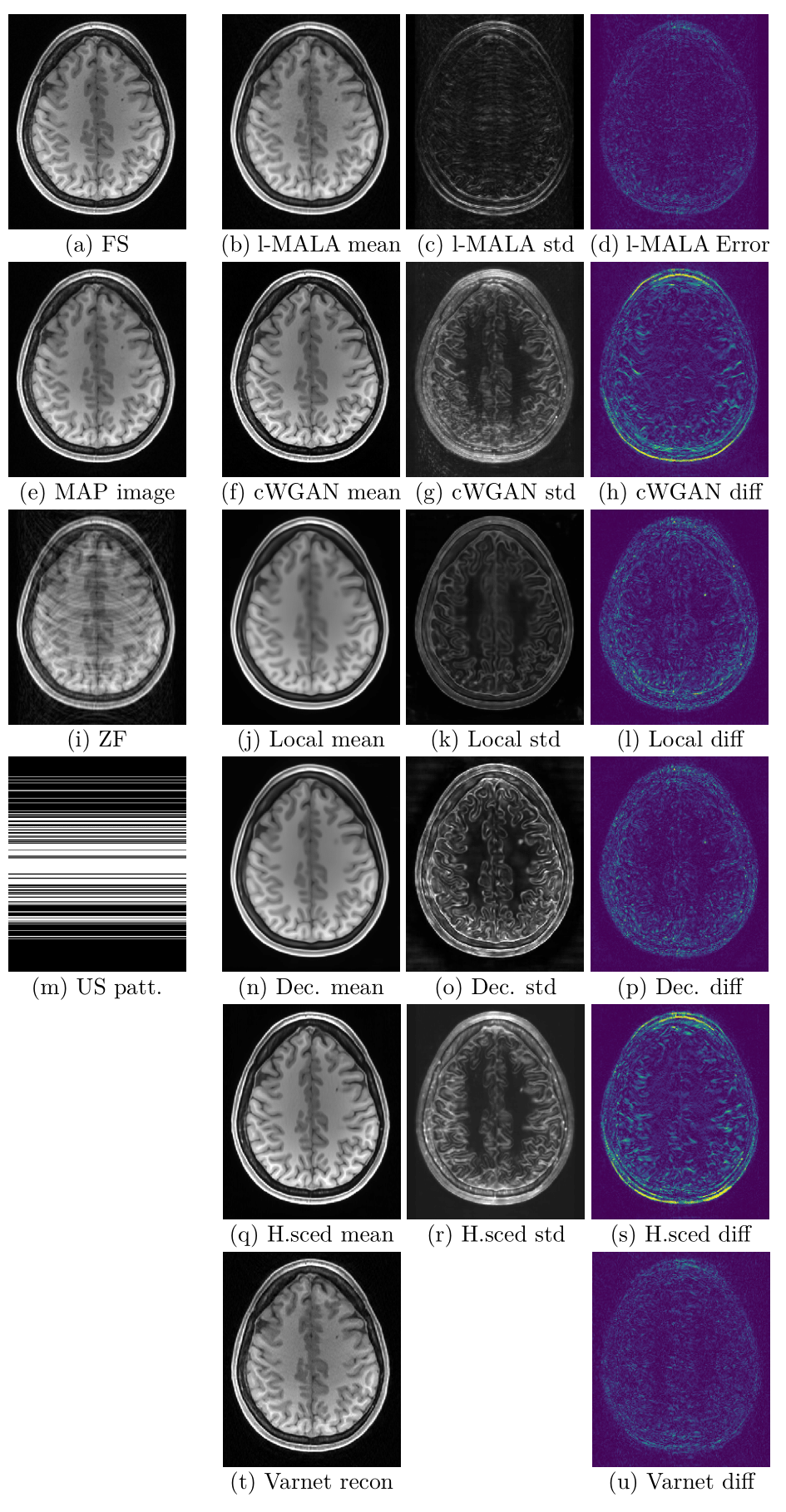}
    \caption{Comparisons at R=3. Figure description same as in main text.}  
    \label{fig:morecomp2}
    \vspace{-0.6cm}
\end{figure}

\begin{figure}[h!]
    \centering
    \includegraphics[width=0.77\textwidth]{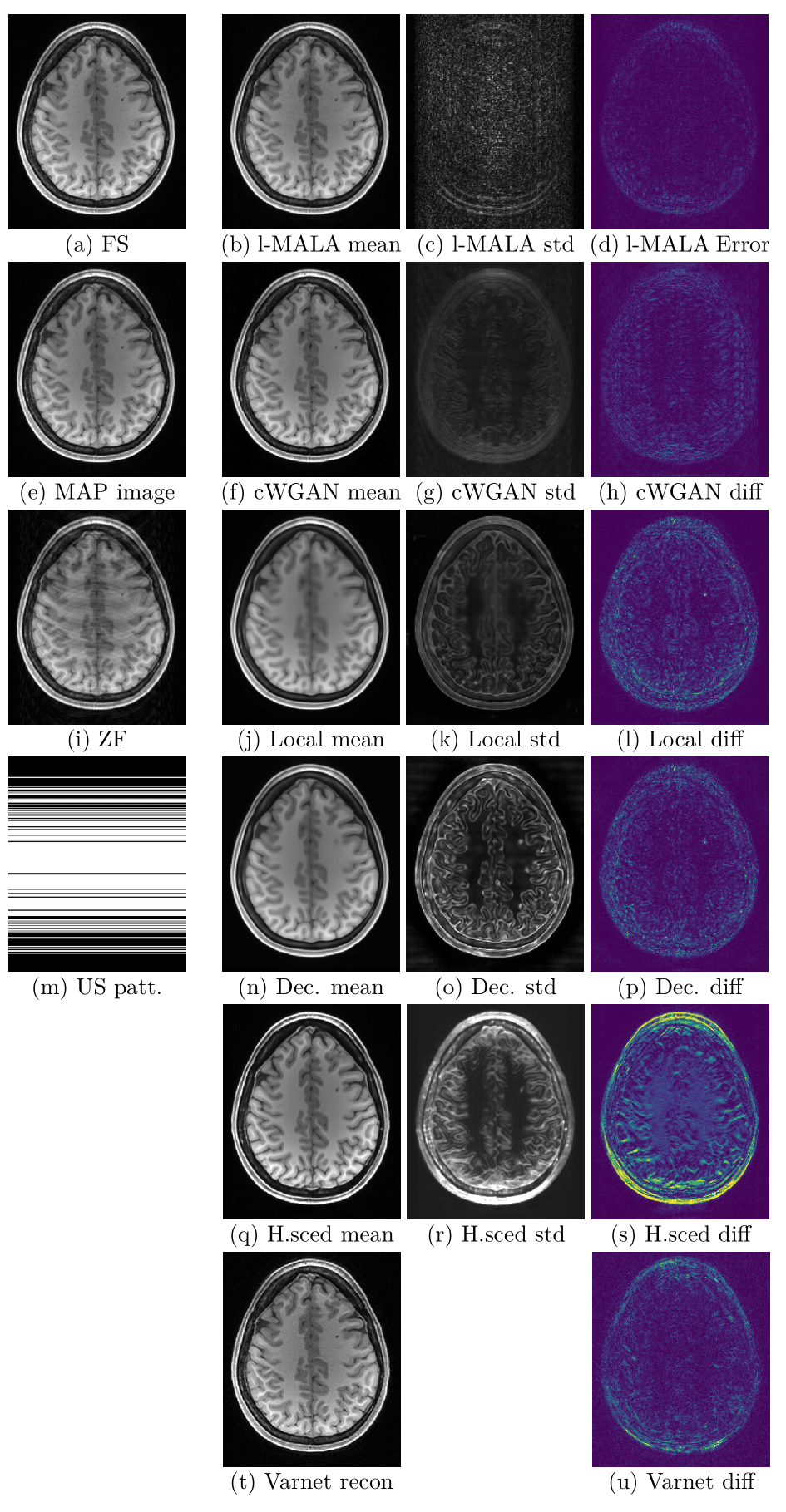}
    \caption{Comparisons at R=2. Figure description same as in main text.}  
    \label{fig:morecomp3}
    \vspace{-0.6cm}
\end{figure}

\begin{figure}[h!]
    \centering
    \includegraphics[width=0.77\textwidth]{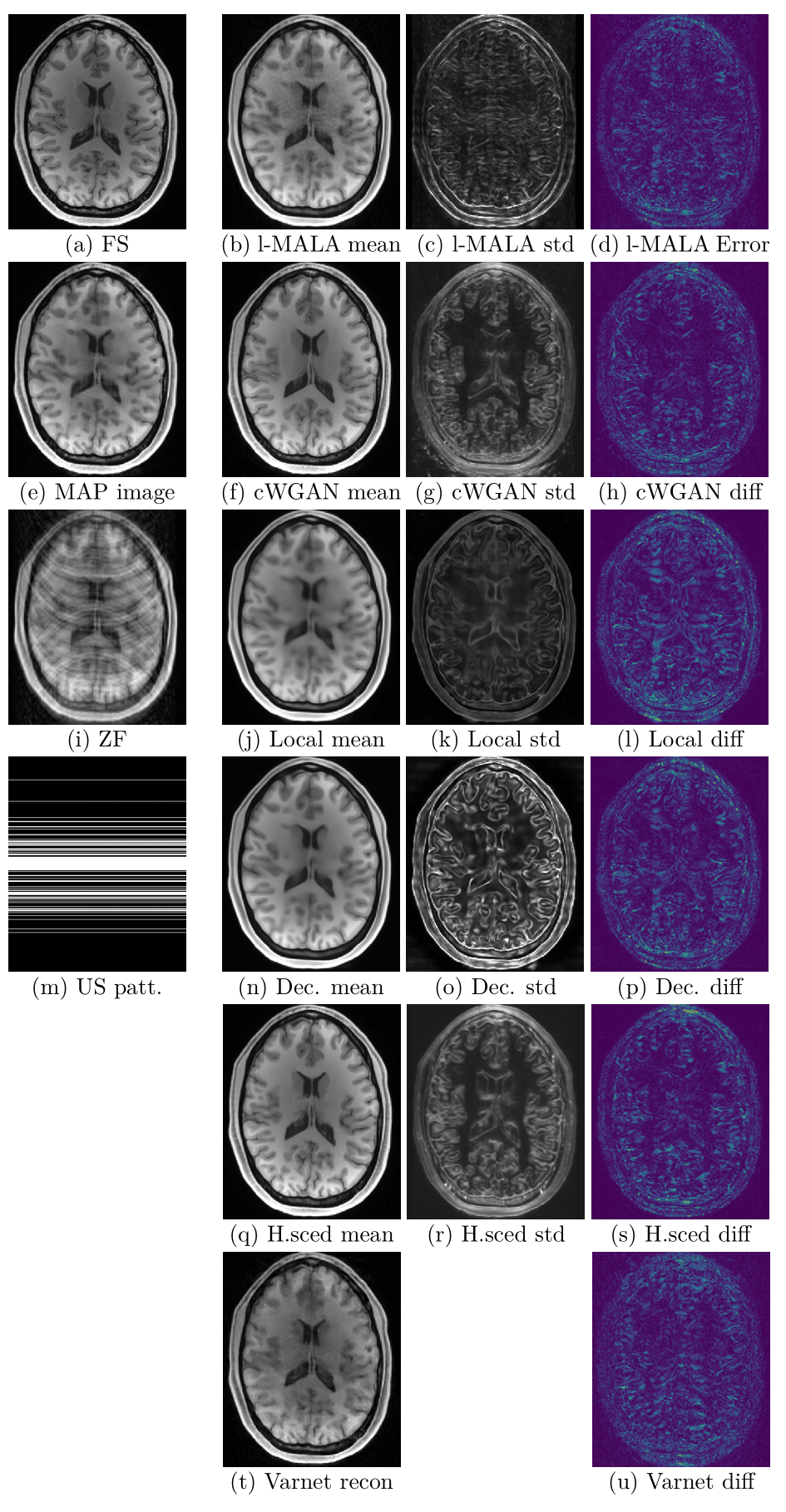}
    \caption{Comparisons at R=5. Figure description same as in main text.}  
    \label{fig:morecomp4}
    \vspace{-0.6cm}
\end{figure}

\begin{figure}[h!]
    \centering
    \includegraphics[width=0.77\textwidth]{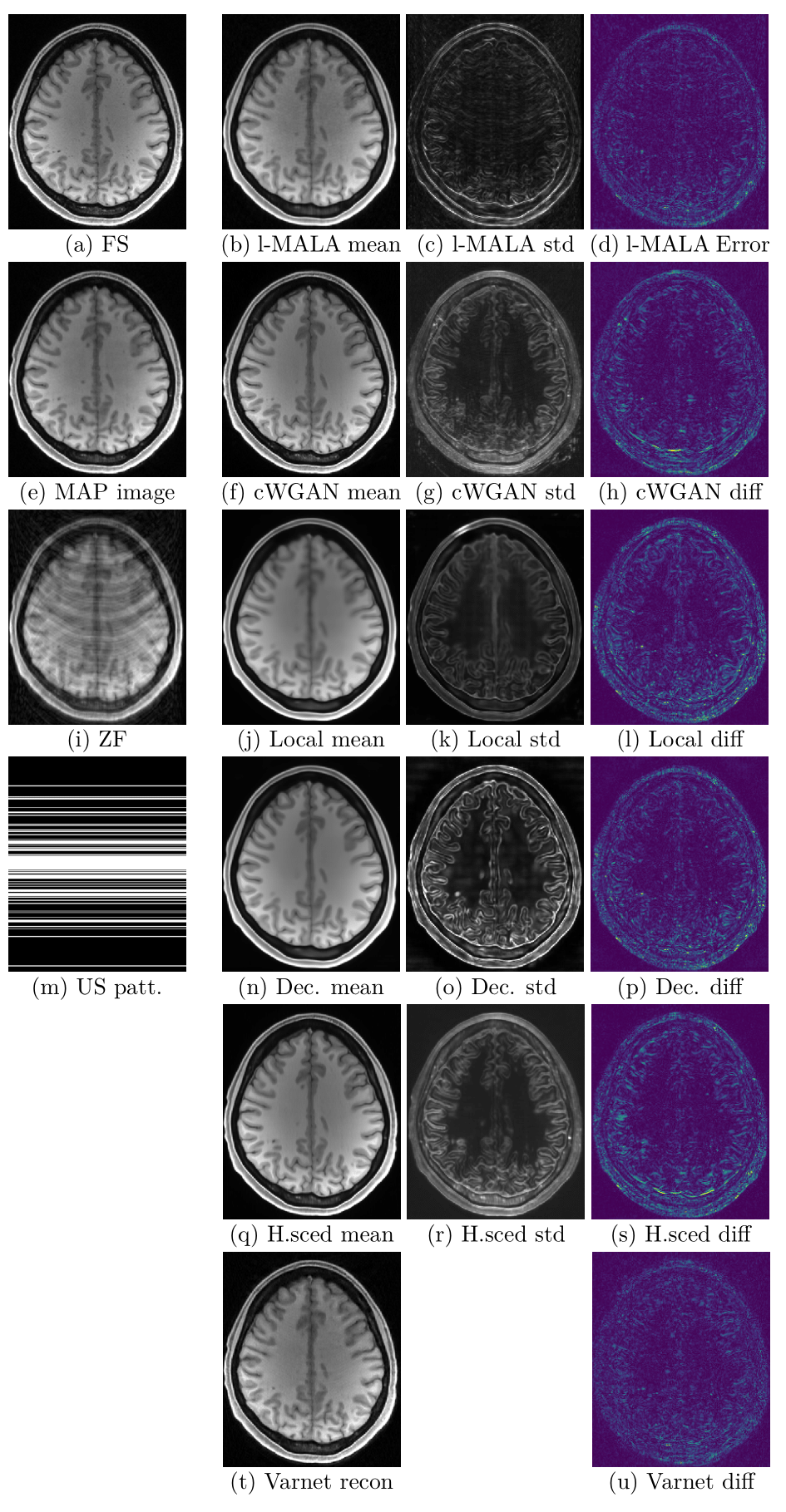}
    \caption{Comparisons at R=4. Figure description same as in main text.}  
    \label{fig:morecomp5}
    \vspace{-0.6cm}
\end{figure}

\begin{figure}[h!]
    \centering
    \includegraphics[width=0.77\textwidth]{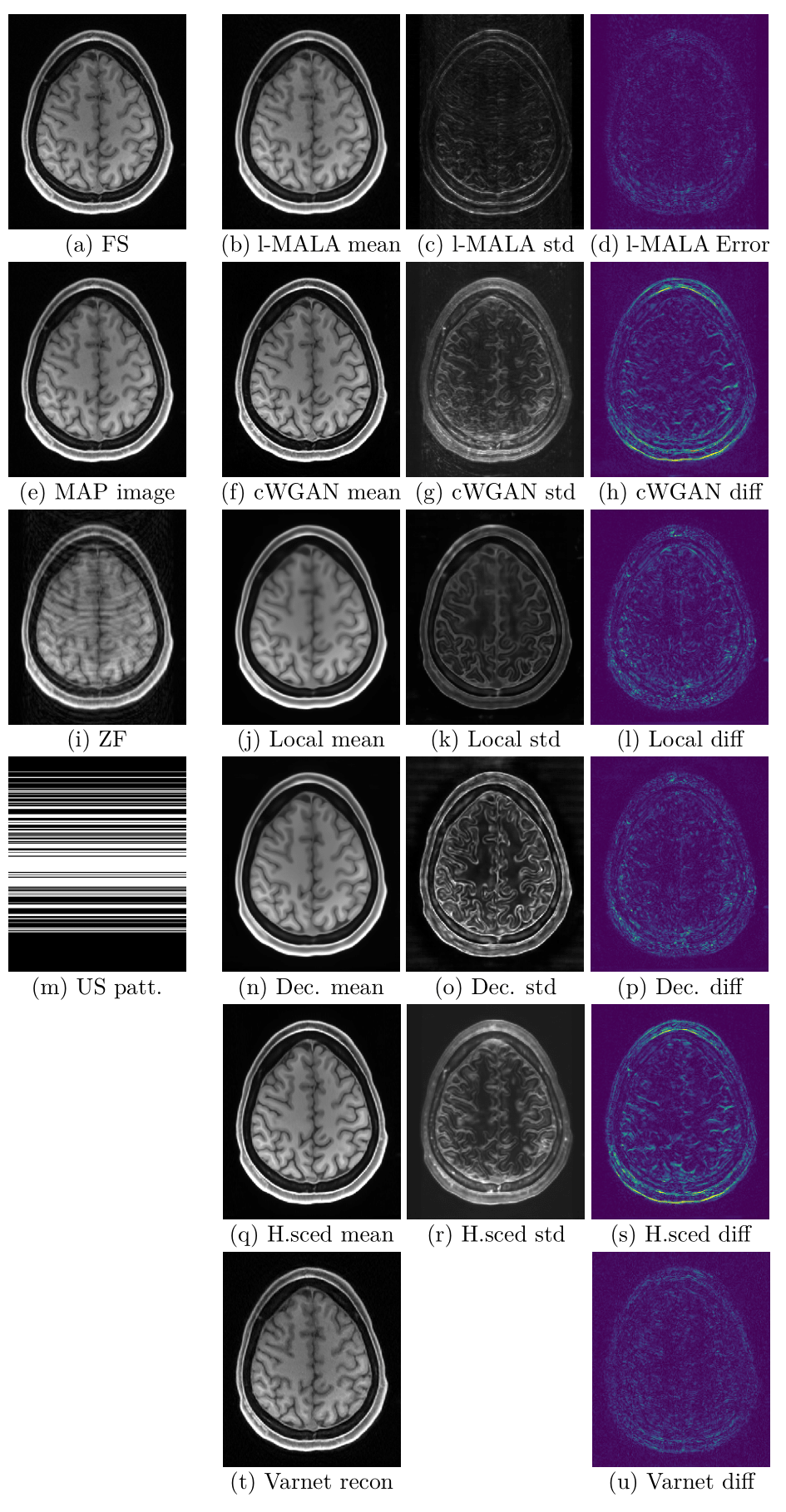}
    \caption{Comparisons at R=3. Figure description same as in main text.}  
    \label{fig:morecomp6}
    \vspace{-0.6cm}
\end{figure}

\begin{figure}[h!]
    \centering
    \includegraphics[width=0.77\textwidth]{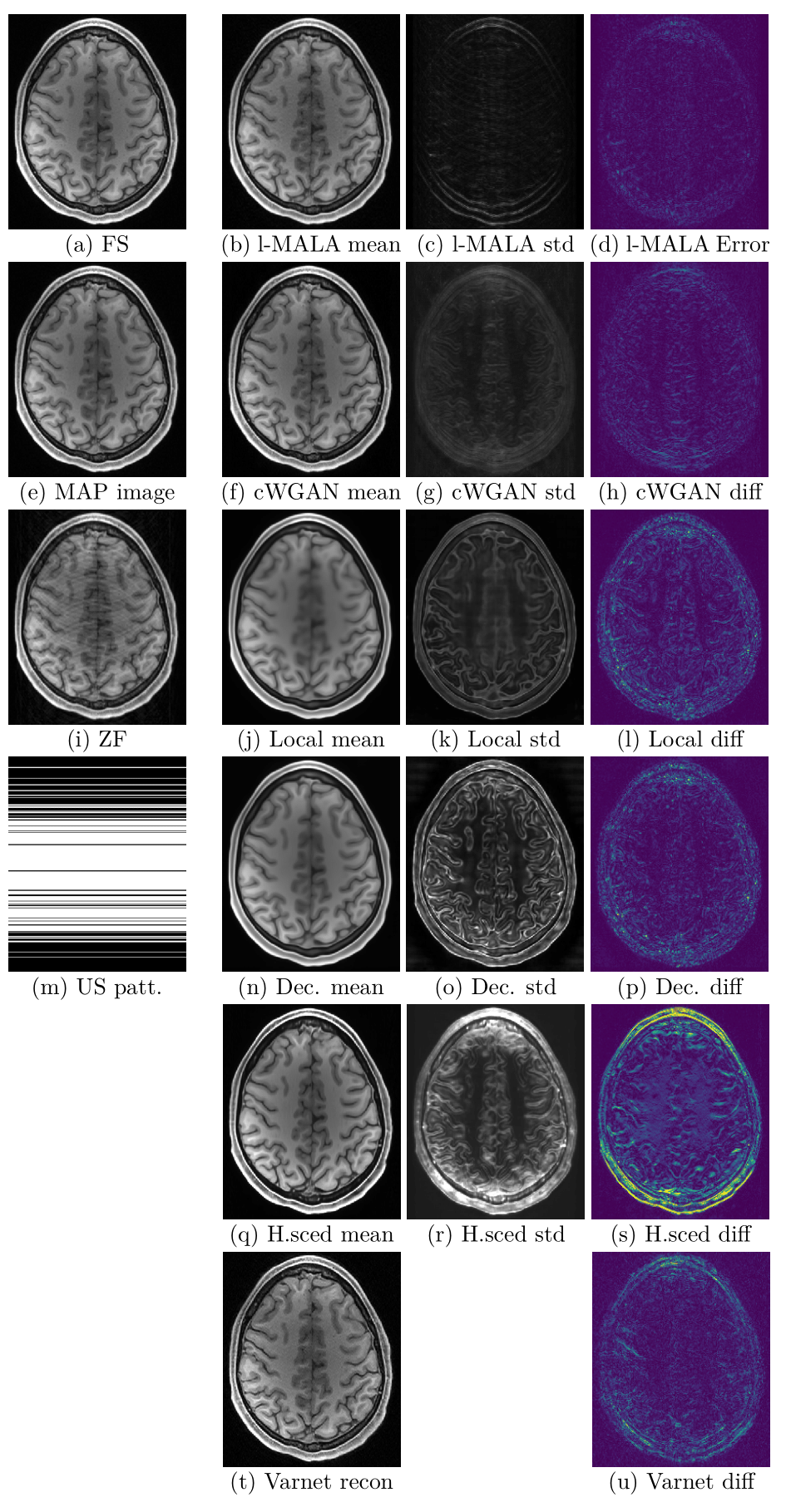}
    \caption{Comparisons at R=2. Figure description same as in main text.}  
    \label{fig:morecomp7}
    
\end{figure}

}

\end{document}